\documentclass{article}
\usepackage{arxiv}
\usepackage[utf8]{inputenc} 
\usepackage[T1]{fontenc}    
\usepackage{hyperref}      
\usepackage{url}           
\usepackage{booktabs}       
\usepackage{amsfonts}      
\usepackage{nicefrac}      
\usepackage{microtype}     
\usepackage{graphicx}
\usepackage{cite}
\usepackage{subfig}
\usepackage{float}
\usepackage[export]{adjustbox}
\usepackage{geometry}
\usepackage{color, colortbl}
\usepackage{rotating}
\usepackage{subfig}
\usepackage[all]{xy}
\usepackage{tikz}
\usepackage{xcolor}
\usepackage{animate}
\usetikzlibrary{shapes.geometric, arrows}
\usepackage{amsmath}
\usepackage[ruled,vlined]{algorithm2e}

\newcommand{\vect}[1]{\boldsymbol{#1}}

\newcommand{\black}[1]{\textcolor[rgb]{0,0,0}{#1}}
\newcommand{\blue}[1]{\textcolor[rgb]{0,0,0}{#1}}

\title{Functional Connectome of the Human Brain with Total Correlation}

\author{
    Qiang Li\\
    Image Processing Laboratory\\
    University of Valencia\\
    Valencia, 46980  \\
  \texttt{qiang.li@uv.es} \\
 \And
    Greg Ver Steeg\\
    Information Sciences Institute\\
    University of Southern California\\
    Marina del Rey, CA 90292\\
    \texttt{gregv@isi.edu}\\
  \And
    Shujian Yu\\
    Machine Learning Group\\
    UiT - The Arctic University of Norway\\
    9037 Troms{\o}, Norway  \\
  \texttt{shujian.yu@uit.no} \\
  \And
    Jesus Malo\\
    Image Processing Laboratory\\
    University of Valencia\\
    Valencia, 46980  \\
}

\begin{document}
\definecolor {processblue}{cmyk}{0.96,0,0,0}
\maketitle

\begin{abstract}

Recent studies proposed the use of Total Correlation to describe functional 
connectivity among brain regions as a multivariate alternative to conventional pair-wise measures such as correlation or mutual information. 
In this work we build on this idea to infer a large scale (whole brain) connectivity network based on Total Correlation and show the possibility of using this kind of networks as biomarkers of brain alterations. 
In particular, this work uses Correlation Explanation (CorEx) to estimate Total Correlation. 
First, we prove that CorEx estimates of total correlation and clustering results are trustable compared to ground truth values. 
Second, the inferred large scale connectivity network extracted from the more extensive open fMRI datasets is consistent with existing neuroscience studies but, interestingly, can estimate additional relations beyond pair-wise regions. 
And finally, we show how the connectivity graphs based on Total Correlation can also be an effective tool to aid in the discovery of brain diseases.

\end{abstract}

\section{Introduction}

%
%
%
%
%
%

\blue{The human brain is a complex system comprised of interconnected functional units.} Millions of neurons in the brain interact with each other at both a structural and functional level to drive efficient inference and processing in the brain. Furthermore, the functional connectivity among these regions also reveals how they interact with each other in specific cognitive tasks. Functional connectivity refers to the statistical dependency of activation patterns between various brain regions that emerges as a result of direct and indirect interactions~\cite{Friston11,Porta14}. \blue{It is usually measured by how similar neural time series are to each other, and it shows how the time series statistically interact with each other.}

A variety of ways to analyze functional connectivity exist. A seed-wise analysis can be performed by selecting a seed-driven hypothesis and analyzing its statistical dependencies with all other voxels outside its limits. \blue{It's a common tool for studying how different parts of the brain are connected to one another. Connectivity is determined by calculating the correlation between the time series of each voxel in the brain and the time series of single seed voxel.} Another option is to perform a wide analysis of the voxel or region of interest (ROI), where statistical dependencies on all voxels or ROIs are studied~\cite{Heuvel10}. Structural connectivity refers to the anatomical organization of the brain by means of fiber tracts~\cite{Sporns05}. The sharing of communication between neurons in multiple regions is coordinated dynamically via changes in neural oscillation synchronizations~\cite{Bastos16}. When it comes to the brain connectome, functional connectivity refers to how different areas of the brain communicate with one another during task-related or resting-state activities~\cite{Lizier11}. \blue{The use of information-theoretic metrics can efficiently detect their interaction in dynamical brain networks, and it's widely used in the field of neuroscience~\cite{Piasini19}. For instance, quantify information encoding and decoding in the neural system~\cite{Ince16, Dimitrov11, Borst99, Bialek12}, measure visual information flow in the biological neural networks~\cite{Gomez19, Malo20}, and color information processing in the neural cortex~\cite{Malo22}, and so on. However, although functional connectivity has already become a hot research topic in neuroscience~\cite{Farahani19,Sporns18}, systematic studies on the information flow or the redundancy and synergy amongst brain regions remain limited. One extreme type of redundancy is full synchronization, where the state of one neural signal may be used to predict the status of any other neural signal, and this concept of redundancy is thus viewed as an extension of the standard notion of correlation to more than two variables~\cite{Rosas18}. Synergy, on the other hand, is analogous to those statistical correlations that govern the whole but not its constituent components~\cite{Rosas19}. High-order brain functions are assumed to require synergies, which give simultaneous local independence and global cohesion, but are less suitable for them under high synchronization situations, such as epileptic seizures~\cite{Tononi99}.} Most functional connectivity approaches until now have mainly concentrated on pairwise relationships between two regions. 
The conventional approach used to estimate indirect functional connectivity among brain regions is Pearson correlation \black{(CC)}~\cite{Pereda05} and Mutual Information (I)~\cite{Chai09, Wang15, Jomaa19, Ince16}.
However, real brain network relationships are often complex, involving more than two regions, and the pairwise dependencies measured by correlation or mutual information could not reflect these multivariate dependencies. Therefore, recent studies in neuroscience focus on the development of information-theoretic measures that can handle more than two regions simultaneously such as the Total Correlation~\cite{Li22b,Li22a}.


Total Correlation (TC)~\cite{Watanabe60} (also known as multi-information~\cite{Studeny98,Laparra11,laparra2020}) mainly describes the amount of dependence observed in the data and, by definition can be applied to multiple multivariate variables. Its use to describe functional connectivity in the brain was first proposed as a empirical measure in~\cite{Li22b}, but in~\cite{Li22a} the superiority of TC over mutual information was proved analytically. The consideration of low-level vision models allows to derive analytical expressions for the TC as a function of the connectivity. These analytical results show that pairwise I cannot capture the effect of different intra-cortical inhibitory connections while the TC can.
Similarly, in analytical models with feedback, synergy can be shown using TC, while it is not so obvious using mutual information~\cite{Li22a}.
Moreover, these analytical results allow to calibrate computational estimators of TC.



In this work we build on these empirical and theoretical results~\cite{Li22a,Li22b} to infer a larger scale (whole brain) network based on TC for the first time.
As opposed to~\cite{Li22a,Li22b} where the number of considered nodes was limited to the range of tens and focused on specialized subsystems, here we consider wider recordings~\cite{Van13,Essen12} so we use signals coming from hundreds of nodes across the whole brain. Additionally, applying our analysis to data of the same scale for regular and altered brains\footnote{\url{http://fcon_1000.projects.nitrc.org/indi/ACPI/html/}}. We also show the possibility of using this kind of wide-range networks as biomarkers. From the technical point of view, here we use Correlation Explanation (CorEx)~\cite{steeg2014NIPS, steeg2015corex_theory} to estimate TC in these high-dimensional scenarios.
Furthermore, graph theory and clustering~\cite{Farahani19,Sporns18} are used here to represent the relationships between the considered regions.


The rest of this paper is organized as follows: Section 2 introduces the necessary information-theoretic concepts and explains CorEx. Sections 3 and 4 show two synthetic experiments that prove that CorEx results are trustable. Section 5 estimates the large-scale connectomes with fMRI datasets that involve more than 100 regions across the whole brain. Moreover, we show how the analysis of these large scale networks based on TC may indicate brain alterations. Sections 6 and 7 give a general discussion and the conclusion of the paper, respectively.

\section{Total Correlation as neural connectivity descriptor}

\subsection{Definitions and Preliminaries}

\textbf{Mutual Information:} Given two multivariate random variables $X_1$ and $X_2$, the mutual information between them, $\mathrm{I}(X_1 ; X_2)$, can be calculated as the difference between the sum of individual entropies, $H(X_i)$ and the entropy of the variables considered jointly as a single system, $H(X_1,X_2)$~\cite{Cover06}:
\begin{equation}
\mathrm{I}(X_1 ; X_2)=\mathrm{H}(X_1)+\mathrm{H}(X_2)-\mathrm{H}(X_1, X_2)
\label{defI}
\end{equation}

where for each (multivariate) random variable $\mathrm{v}$, the entropy is $\mathrm{H}(\mathrm{v})=\left\langle-\log _{2} \mathrm{p}(\mathrm{v})\right\rangle$ and the brackets represent expectation values spanning random variables.
The mutual information also can be seen as the information shared by the two variables or the reduction of uncertainty in one variable given the information about the other~\cite{Kraskov04}.

\textbf{Mutual information is better than linear correlation:} For Gaussian sources mutual information reduces to linear correlation because the entropy factors in Eq.~\ref{defI} just depend on $|\langle X_1 \cdot X_2^\top \rangle|$. However, for more general (non-Gaussian) sources mutual information cannot be reduced to covariance and cross-covariance matrices. In these (more realistic) situations I is better than the linear correlation because I captures nonlinear  relations that are ruled out by $|\langle X_1 \cdot X_2^\top \rangle|$. 
For an illustration of the qualitative differences between I and linear correlation see the examples in Section 2.2 of~\cite{Li22b}.

As a result, mutual information has been proposed as a good alternative to linear correlation for estimating functional connectivity~\cite{Chai09, Ince16}. However, mutual information cannot not capture dependencies beyond pairs of nodes. And this may be a limitation in complex networks~\cite{Fabila22}.

\begin{figure}[t]
    \centering
    \includegraphics[width=\textwidth, height=13em, center]{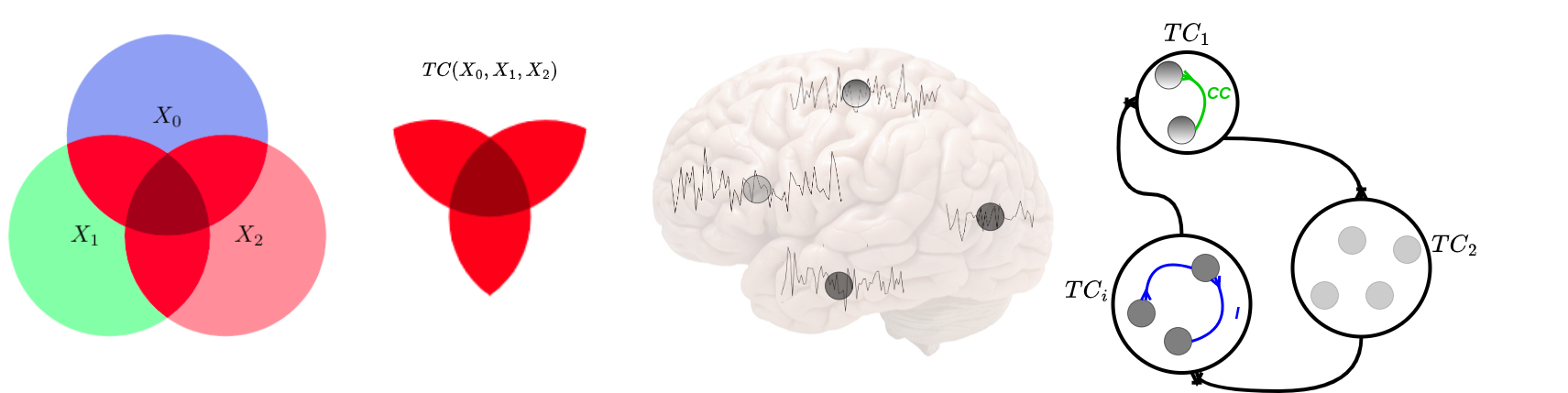}
    \caption{\textbf{Conceptual scheme of information theoretic measures of neural information flow.} The left circle areas represent amounts of information, and intersections represent shared information among the corresponding variables, $X_{0},X_{1},X_{2}$. Examples of entropy, $H(X_{0}),H(X_{1}),H(X_{2})$, total correlation (red color), and $TC[X_{0},X_{1},X_{2}]$ are given. \blue{The middle figures show some neural time series are extracted from brain regions, which correspond to the nodes in the right figure.}
    The right figures illustrate large-scale time series in the brain and how the coupled information is transmitted among the brain regions. The blue and green lines show linear correlation (CC) and mutual information (I), respectively, between different parts of the brain. The modules represent the lobes of the human brain. Each module has specific brain regions, and each module works with the others.}
    \label{fig:tc_vd}
\end{figure}

\textbf{Total Correlation:} 
This magnitude describes the dependence among $n$ variables and it is a generalization of the mutual information concept from two parties to $n$ parties. The Venn Diagram in Fig.~\ref{fig:tc_vd} qualitatively illustrates this for three variables. The definition of total correlation from Watanabe~\cite{Watanabe60} can be denoted as,

\begin{equation}
\mathrm{T} \mathrm{C}\left(\mathrm{X}_{1}, \ldots, \mathrm{X}_{\mathrm{n}}\right) \equiv \sum_{i=1}^{n} \mathrm{H}\left(\mathrm{X}_{\mathrm{i}}\right)-\mathrm{H}\left(\mathrm{X}_{1}, \ldots, \mathrm{X}_{\mathrm{n}}\right)={\mathrm{D}}_{\mathrm{KL}}\left(\mathrm{p}\left(\mathrm{X}_{1}, \ldots, \mathrm{X}_{\mathrm{n}}\right) \| \prod_{\mathrm{i}=1}^{\mathrm{n}} \mathrm{p}\left(\mathrm{X}_{\mathrm{i}}\right)\right)
\label{defT}
\end{equation}

where $\mathrm{X} \equiv\left(\mathrm{X}_{1}, \ldots, \mathrm{X}_{\mathrm{n}}\right)$, and TC can also be expressed as the Kullback Leibler divergence, $D_{KL}$ between the joint probability density and the product of the marginal densities. 
From these definitions, if all variables are independent then TC will be zero.

\blue{The conditional total correlation, which is similar to the definition of total correlation but with a condition appended to each term, The Kullback-Leibler divergence of the two conditional probability distributions can also be used to define the conditional total correlation.} The estimation method used in this work (CorEx presented in the next subsection) uses the TC after conditioning on some other variable $Y$, which can be defined as~\cite{Cover06},

\begin{equation}
    TC(X|Y) =\sum_{i}H(X_{i}|Y)-H(X|Y) = {D}_{KL}(p(\mathrm{x}|\mathrm{y}) \| \prod_{\mathrm{i}=1}^{\mathrm{n}} \mathrm{p}(\mathrm{x}_{\mathrm{i}}|\mathrm{y}))
\end{equation}

\textbf{Total correlation is better than Mutual information:} 
This superiority is not only due to the obvious $n$-wise versus pair-wise definitions in Eqs.~\ref{defI} and~\ref{defT}. It also has to do with the different properties of these magnitudes.
To illustrate this point let us recall one of the analytical examples in~\cite{Li22a}. Consider the following feedforward network:
\begin{equation}
\xymatrix{X_1 \ar[r]^{} &  X_2 \ar[r]^{}  & \mathbf{e} \ar[r]^{f} & X_3}
\label{Framework}
\end{equation}
\blue{where the nodes $X_{1}$, $X_{2}$, $e$, and $X_{3}$ can have any number of neurons, the first two transforms, $\xymatrix{X_1 \ar[r]^{} &  X_2 \ar[r]^{}  & \mathbf{e}}$, are linear and affected by additive noise, and the last transform, $f(\cdot)$, is nonlinear but deterministic.
Imagine that in this network one is interested in the connectivity between the neurons in the hidden layer, $\mathbf{e}$, but the nonlinear function~$f(\cdot)$ is unknown and one only has experimental access to the signal in the regions $X_1$, $X_2$ and $X_3$.}
%
In this situation one could think on measuring $I(X_1,X_3)=I(X_1,f(\mathbf{e}))$ or $I(X_2,X_3)=I(X_1,f(\mathbf{e}))$. However, the invariance of $I$ under arbitrary nonlinear re-parametrization of the variables~\cite{Kraskov04} implies that these measures are insensitive to $f$ and the connectivity there in. 
On the contrary, as pointed out in~\cite{Li22a}, using the expression for the variation of TC under nonlinear transforms~\cite{Lyu09,Malo20}, the variation of $H$ under nonlinear transforms~\cite{Cover06}, and the definition in Eq.~\ref{defT}, one obtains $TC(X_1,X_2,X_3) = [TC(X_1,X_2,\mathbf{e})-TC(\mathbf{e})]+TC(X_3)$, where the term in the bracket does not depend on $f(\cdot)$, but the last term definitely does, which proves the superiority of $TC$ over $I$ in describing connectivity. 

\blue{In~\cite{Li22a} the network in Eq.~\ref{Framework} specifically refers to the flow from the retina, $X_1$, to the LGN, $X_2$, and finally to the visual cortex, $\vect{e}$ and $X_3$. However, the result of the superiority of $TC$ over $I$ to describe the connectivity in the hidden layer is totally general for every network with the generic properties listed after Eq.~\ref{Framework}.}

\subsection{Total Correlation estimated from CorEx}

Straightforward application of the direct definition of TC is not feasible in high dimensional scenarios, and alternatives are required~\cite{Laparra11,laparra2020}.
A practical approach to estimate total correlation is via \emph{latent factor modelling}. \blue{A latent factor model is a statistical model that relates a set of observable variables to a set of latent variables. The idea is to explicitly construct latent factors, $Y$, that somehow capture the dependencies in the data.} If we measure dependencies via total correlation, $TC(X)$, then we say that the latent factors \emph{explain} the dependencies if $TC(X|Y)=0$. We can measure the extent to which $Y$ explains the correlations in $X$ by looking at how much the total correlation is reduced
\begin{align}
\label{eq:tcbound}
TC(X) - TC(X|Y) = \sum_{i=1}^n I(X_i;Y) - I(X;Y)
\end{align}
The total correlation is always non-negative, and the decomposition on the right in terms of mutual information can be verified directly from the definitions. Any latent factor model can be used to lower bound total correlation, and the terms on the right-hand side of Eq. \ref{eq:tcbound} can be further lower-bounded with tractable estimators using variational methods, and Variational Autoencoders (VAEs) are a popular example \cite{aistats2019}.  

\blue{Although latent factor models do not give a direct total correlation estimation as the Rotation-based Iterative Gaussianization (RBIG)~\cite{Laparra11,laparra2020} and the Matrix-based R\'enyi's entropy~\cite{yu2019multivariate} did, the approach can be complementary because the construction of latent factors can help in dealing with the curse of dimensionality and for interpreting the dependencies in the data. Compared to CorEx, the main goal of RBIG\footnote{\black{https://isp.uv.es/RBIG4IT.htm}} is to convert any non-Gaussian distribution data into a Gaussian distribution through marginal Gaussization and rotation to get TC. The Matrix-based R\'enyi's entropy\footnote{http://www.cnel.ufl.edu/people/people.php?name=shujian} is mainly used for estimating multivariate information based on Shannon’s entropy, which is R\'enyi's $\alpha-$order entropy~\cite{Shannon48}. With these goals in mind, we now describe a particular latent factor approach known as Total Cor-relation Ex-planation (CorEx\footnote{https://github.com/gregversteeg/CorEx})\cite{steeg2014NIPS}.}

CorEx constructs a factor model by reconstructing latent factors using a factorized probabilistic function of the input data, $p(y|x) = \prod_{j=1}^m p(y_j|x)$, with $m$ discrete latent factors, $Y_j$. This function is optimized to give the tightest lower bound possible for Eq.~\ref{eq:tcbound}.  
\begin{align}
TC(X) \geq \max_{p(Y_j|x)} \sum_{i=1}^n I(X_i;Y)  - I(X;Y) = \sum_{j=1}^m \left(\sum_{i=1}^n \alpha_{i,j} I(X_i;Y_j) -  I(Y_j;X)\right)
\end{align}
The factorization of the latent leads to the terms $ I(X;Y) = \sum_j I(Y_j;X)$ which can be directly calculated. The term $I(X_i;Y)$ is still intractable and is decomposed using the chain rule into $I(X_i;Y) \approx \sum \alpha_{i,j} I(X_i;Y_j) $. Each $I(X_i;Y_j)$ can then be tractably estimated~\cite{steeg2015corex_theory,steeg2014NIPS}. There are free parameters $\alpha_{i,j}$ that must be updated while searching for latent factors and achieving objective functions. When $t=0$, the $\alpha_{i,j}$ initializes and then updates according to:

\begin{equation}
    \alpha_{i, j}^{t+1}=(1-\lambda) \alpha_{i, j}^t+\lambda \alpha_{i, j}^{* *}
\end{equation}

The second term $\alpha_{i, j}^{* *}=\exp \left(\gamma\left(I\left(X_i: Y_j\right)-\max _j I\left(X_i: Y_j\right)\right)\right)$, $\lambda$ and $\gamma$ are constant parameters. This decomposition allows us to quantify the contribution to the total correlation bound from each latent factor, which we can aid interpretability. 

CorEx can be further extended into a hierarchy of latent factors \cite{steeg2015corex_theory}, helping to reveal hierarchical structure that we expect to play an important role in the brain. The latent factors at layer $k$ explain dependence of the variables in the layer below. 

\begin{equation}
TC(X) \geq \sum_{k=1}^r \left( \sum_{j=1}^m \left(\sum_{i=1}^n \alpha_{i,j}^k I(Y_i^{k-1};Y_j^k) - \sum_{j=1}^m I(Y_j^k;Y^{k-1}) \right)\right)
\end{equation}

\begin{figure}[htbp]
    \centering
    \includegraphics[width=0.76\textwidth]{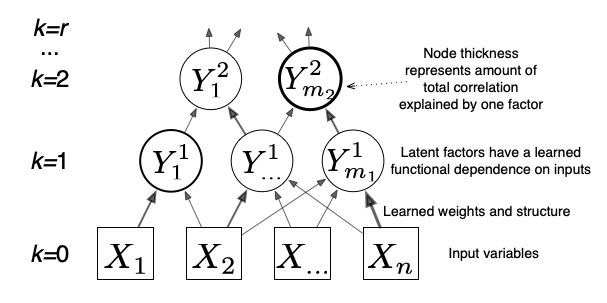}
    \caption{\textbf{CorEx learns a hierarchical latent factor as illustrated above.} Edge thickness indicates strength of relationship between factors, and node thickness indicates how much total correlation is explained by each latent factor.}
    \label{fig:corex}
\end{figure}

Here $k$ gives the layer and $Y^0 \equiv X$ denotes the observed variables. Ultimately, we have a bound on TC that gets tighter as we add more latent factors and layers and for which we can quantify the contribution for each factor to the bound. We exploit this decomposition for interpretability~\cite{steeg2017unsupervised} as illustrated in Fig.~\ref{fig:corex}. CorEx prefers to find modular or tree-like latent factor models which are beneficial for dealing with the curse of dimensionality~\cite{ver2019fast}. For neuroimaging, we expect this modular decomposition to be effective because functional specialization in the brain are often associated with spatially localized regions. We explore this hypothesis in the experiments.  

\section{Experiment 1: Total Correlation for independent mixtures}

\begin{figure}[ht!]
    \centering
    \includegraphics[width=\textwidth, height=15em]{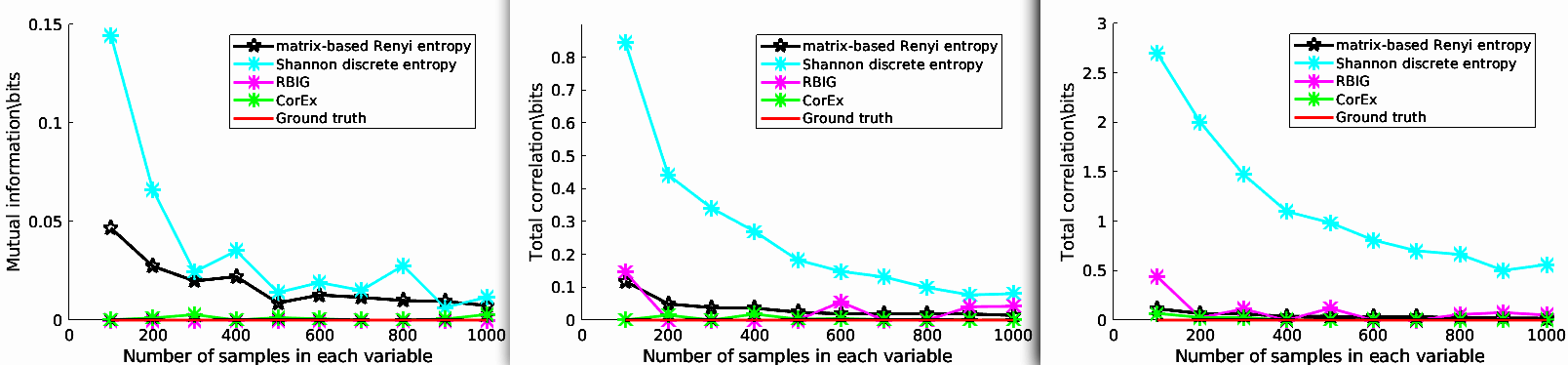}
    \caption{\textbf{The estimated total correlation values for three independent variables.} The various total correlation estimators are compared with ground truth value (red line), for example, matrix-based Renyi entropy (black line), Shannon discrete entropy (cyan line), RBIG (\black{magenta} line), and CorEx (green line). See the main text for more information.}
    \label{Fig:MITC_Simu}
\end{figure}

In this experiment, we estimate the total correlation of three independent variables $X$, $Y$ and $Z$, and each follows a Gaussian distribution. For this setup, the ground truth of TC should satisfy $TC(X, Y, Z) = 0$, and generated various samples with different lengths. Then estimated total correlation values are shown in the Fig~\ref{Fig:MITC_Simu}. Here, we compared CorEx with other different total correlation estimators, such as, RBIG~\cite{Laparra11,laparra2020}, Matrix-based R\'enyi's entropy~\cite{yu2019multivariate}, Shannon discrete entropy\footnote{https://github.com/nmtimme/Neuroscience-Information-Theory-Toolbox}, and ground truth. The left figure (2 dimensional) is mutual information, and the middle (3 dimensional) and right figure (4 dimensional) are total correlation. As we mentioned above, the simulation data is totally Gaussian distributed. Therefore, their dependency should be zero. We find that CorEx and RBIG both perform very well and are very stable, and matrix-based Renyi entropy performance becomes more and more nice with increased dimensions, while Shannon discrete entropy becomes more and more accurate with an increase of samples. All these make sense, and it also explains the accuracy of total correlation estimation with CorEx. \blue{Here, compared to other estimators, the main functionality goal of CorEx is to cluster statistical dependency variables based on total correlation. However, other estimators mainly focus on directly getting the total correlation value and do not supply very nice visualization results. The CorEx gives us a nice connection with graph theory to visualize and show their functional relationship.} 

\section{Experiment 2: Clustering by Total Correlation for dependent and independent mixtures}

\blue{To evaluate the performance of CorEx in clustering tasks. The elements in group $X$ include $X1$, $X2$, and $X3$, which satisfy Gaussian distributions and are completely independent from each other and from group $Y$, and variables in group $Y$ include $Y1$, $Y2$ from $Y1$, and $Y3$ from $Y2$, which are connected to each other. Then we compare the CorEx cluster results with pairwise Pearson correlation, pairwise mutual information, and partial correlation, which consider confounding effects to find the groups.}

\begin{figure}[htbp]
    \centering
    \includegraphics[width=\textwidth, height=30em]{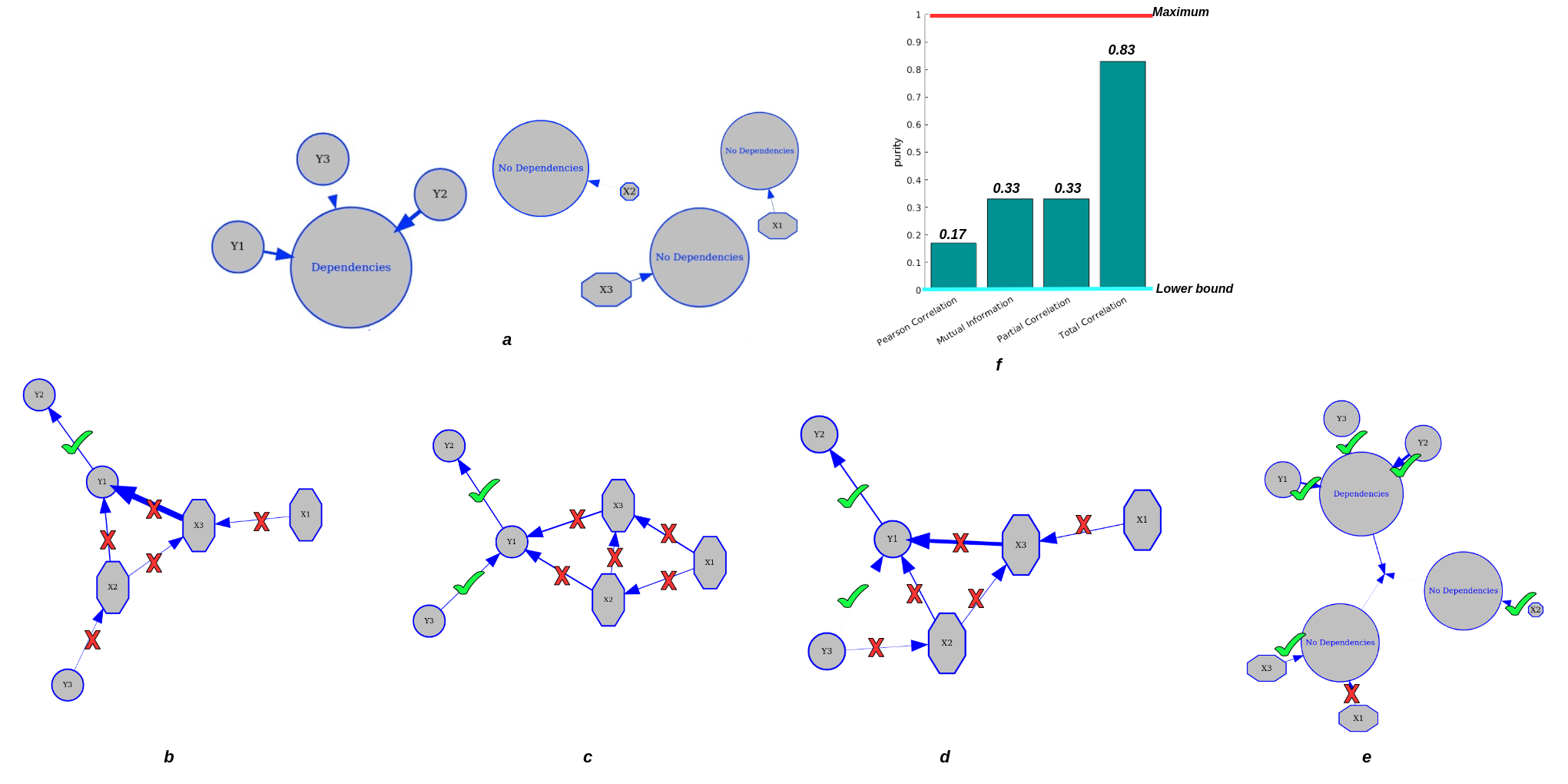}
    \caption{\textbf{Clustering performance for dependent and independent mixtures.} \blue{The top row: \textbf{a} displays the ground truth of variable clustering in two groups. The \textbf{f} shows the purity value of each approach. The second row: \textbf{b} shows the clustering result based on Pearson correlation. The \textbf{c} shows the clustering result by pairwise mutual information. The \textbf{d} shows the clustering result by partial correlation. The \textbf{e} shows clustering results by CorEx based on total correlation.}}
    \label{Fig:clustering_tc}
\end{figure}

In Fig.~\ref{Fig:clustering_tc}, we found that CorEx based on total correlation has high accuracy in estimating their dependencies (Fig.~\ref{Fig:clustering_tc}\textbf{e}) compared to pairwise Pearson correlation (Fig.~\ref{Fig:clustering_tc}\textbf{b}), pairwise mutual information (Fig.~\ref{Fig:clustering_tc}\textbf{c}), and partial correlation (Fig.~\ref{Fig:clustering_tc}\textbf{d}). As we established in this experiment, the elements in group $Y$ should be clustered together, and the elements in group $X$ should be completely independent of each other and of group $Y$. \blue{The ground truth is presented in Fig.~\ref{Fig:clustering_tc}\textbf{a}. Then we estimated the cluster result with pairwise Pearson correlation with threshold $0.1$, pairwise mutual information with threshold $0.4$, and partial correlation without threshold. Obviously, we found that pairwise approaches have high errors in accurately estimating their statistical dependencies, and pairwise mutual information is better than pairwise Pearson correlation, but still has high errors in correctly clustering tasks. When we considered the confounding effect of third variables, we still did not get a better clustering result compared to TC. Therefore, the clustering results with CorEx by total correlation get the best performance compared to pairwise approaches.} Moreover, we use \textit{Purity} as a criterion of clustering quality to qualify the performance of clustering because it's a straightforward and transparent evaluation metric~\cite{manning08}. To calculate purity, each cluster is allocated to the class that occurs most frequently within it, and the accuracy of this assignment is determined by counting the number of correctly assigned elements and dividing by $N(N=6)$. Formally:

\begin{equation}
    \operatorname{Purity}(X, Y)=\frac{1}{N} \sum_i \max _j\left|X_{i} \cap Y_{j}\right|
    \label{Eq.pur}
\end{equation}

\blue{where $X = \{X1, X2, X3 \}$ is the set of clusters and $Y = \{Y1, Y2, Y3 \}$ is the set of classes.} Fig.~\ref{Fig:clustering_tc}\textbf{f} presents the clustering performance of pairwise approaches and CorEx with purity as a criterion. Poor clusters have near-zero purity ratings (lower bound). A perfect cluster possesses a purity of one (maximum value). Based on Eq.\ref{Eq.pur}, we get purity values of $0.17$ and $0.33$ for pairwise approaches and partial correlation, and the purity value for CorEx is $0.83$. All in all, we showed that CorEx based on total correlation has the best performance compared to pairwise approaches.

\section{Experiment 3: Brain functional connectivity analysis using Total Correlation}

\begin{figure}[ht!]
    \centering
    \includegraphics[width=0.7\textwidth]{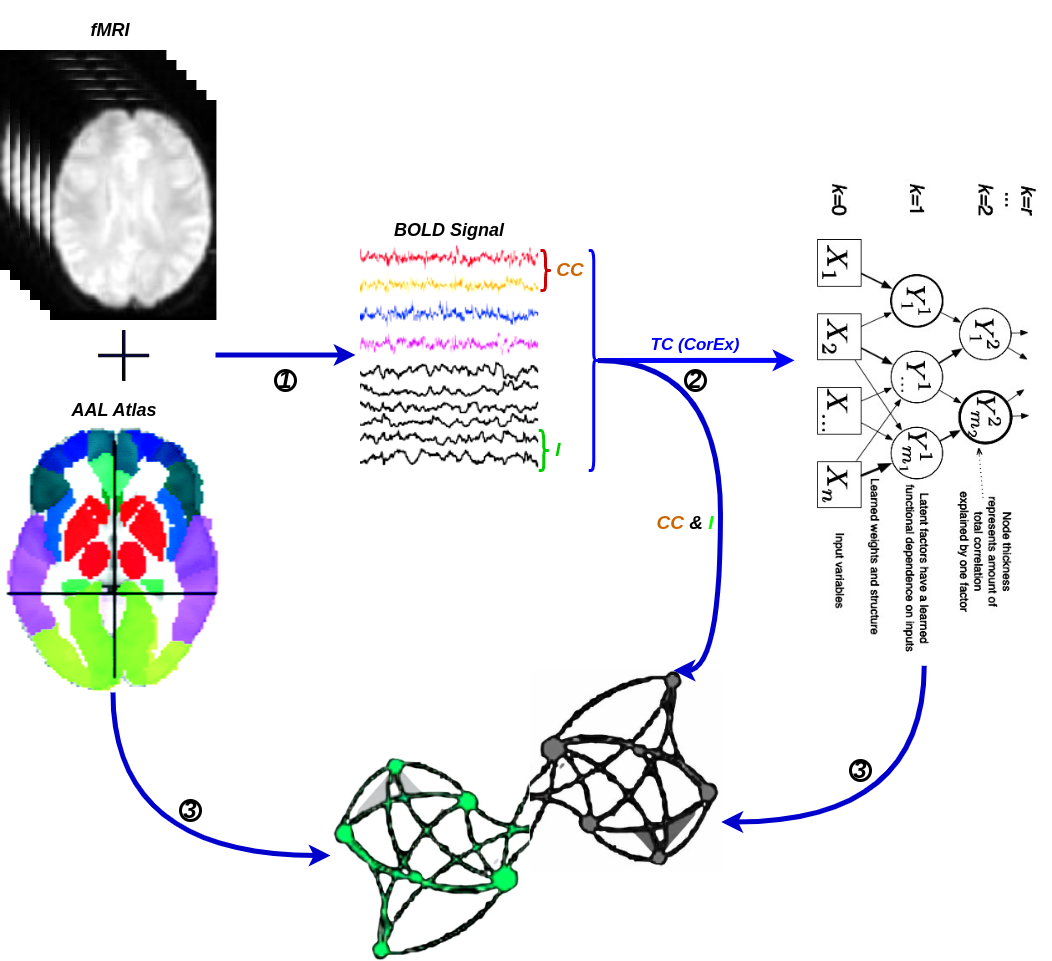}
    \caption{\textbf{A flowchart for the construction of functional brain network by fMRI.} \blue{$\boldsymbol{\textcircled{1}}$ Time series extraction from fMRI data within each anatomical unit (i.e., network node).
    $\boldsymbol{\textcircled{2}}$ Estimation of a functional connectivity with CC, I, and TC (CorEx), respectively. $\boldsymbol{\textcircled{3}}$ Visualization of functional connectivity as a tree and circle graphs (i.e., network edges and network nodes).}}
    \label{Fig:pipeline_fc}
\end{figure}

A network is a collection of nodes and edges, where nodes represent fundamental elements (e.g., brain regions) within the system of interest (e.g., the brain), and edges represent the dependencies that exist between those fundamental elements with considered weights. \blue{Typically, the threshold is chosen based on the visual effect on functional connectivity, and here, we set the optimal threshold for community detection in brain connectivity networks. We use it to identify a threshold that maximizes information on the network modular structure, removes the weakest edges, and keeps the largest connected component. Fig.~\ref{Fig:pipeline_fc} illustrates the schematic representation of network construction using fMRI. Firstly, the time series are extracted from fMRI data based on a selected structural atlas, and then functional connectivity is estimated with CC, I, and CorEx, respectively. The results are presented with a graph that includes both brain nodes and their functional connectivity with weight edges.}

\subsection{First total correlation-based clustering example from fMRI data}

The data was taken from a resting-state fMRI experiment in which a subject was watching and maintaining alert wakefulness but not performing any other behavioral task. Meanwhile, the BOLD signal was recorded. This data was downloaded from Nitime \footnote{\url{https://nipy.org/nitime/index.html}}. The data was preprocessed, and time series were extracted from different regions of interest (ROIs) in the brain. The ROIs abbreviations and related full name were listed as follows: Cau, Caudate; Pau, Paudate; Thal, Thalamus; Fpol, Frontal pole; Ang, Angular gyrus; SupraM, Supramarginal Gyrus; MTG, Middle Temporal Gyrus; Hip, Hippocampus; PostPHG, Posterior; Parahippocamapl gyrus; APHG, Anterior parahippocamapl gyrus; Amy, Amygdala; ParaCing, Paracingulate gyrus; PCC, Posterior cingulate cortex; Prec, Precuneus; R, right hemisphere; L, left hemisphere. First, we estimated the pairwise functional connectivity metrics with Pearson correlation, mutual information, and the corresponding functional connectivity, a circle-weighted graph used to visualize the outcome of pairwise functional connectivity. In Fig.~\ref{Fig:cc_mi_graph} top row (left) and (right), Pearson correlation and mutual information estimate the same pairwise dependencies, but later approaches capture stronger weights between ROIs, such as LPCC and RPCC, LThal and RThal, and LAmy and RAmy.

\begin{figure}[H]
    \centering
    \includegraphics[width=0.9\textwidth, height=11em]{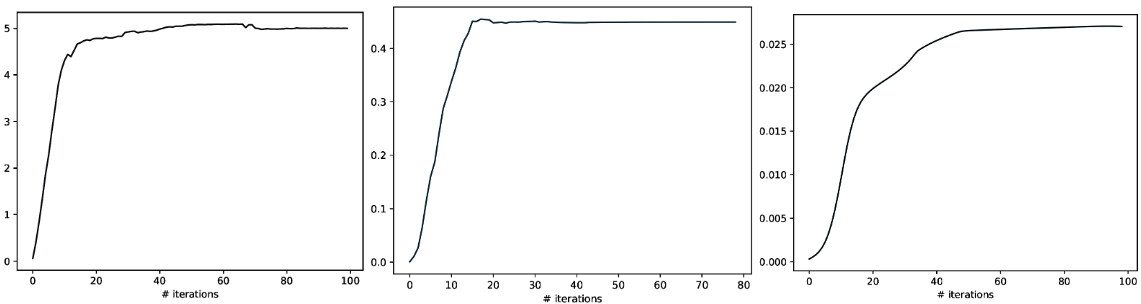}
    \caption{\textbf{The total correlation converge curve of CorEx in layers 1, 2, and 3 is shown above.} From left to right, their corresponding layer1, layer2, and layer3 parameters are selected in event-related experiments, and it shows that the total correlation lower bound stops increasing and tends to converge.}
    \label{Fig:tc_bound}
\end{figure}

\begin{figure}[htbp]
    \centering
    \subfloat[]{\scalebox{1}{\includegraphics[width=1.3\textwidth, height=11cm, center]{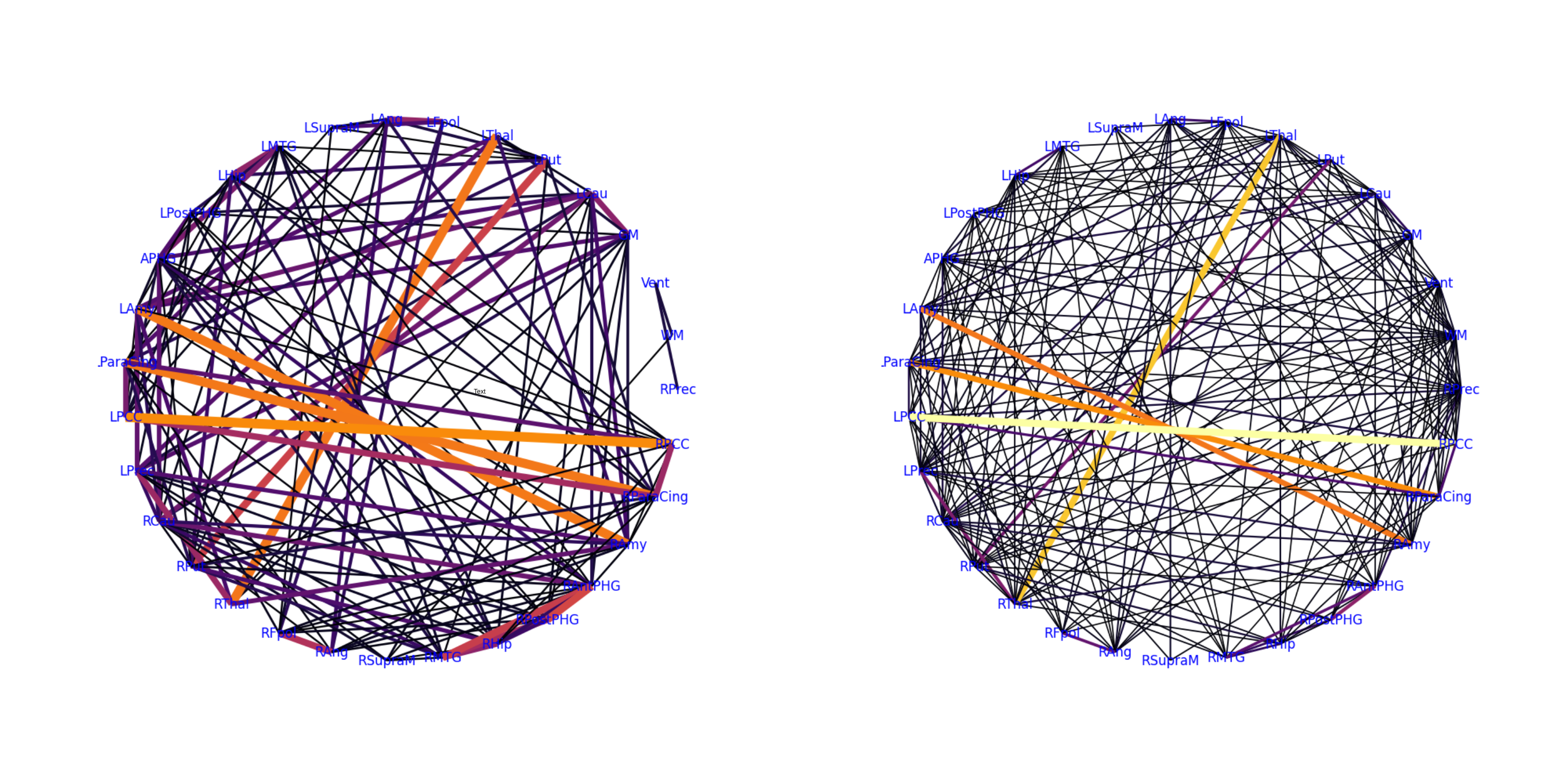}}} \\
    \vspace{2mm}
    \subfloat[]{\scalebox{1}{\includegraphics[width=1.2\textwidth, height=8.5cm, center]{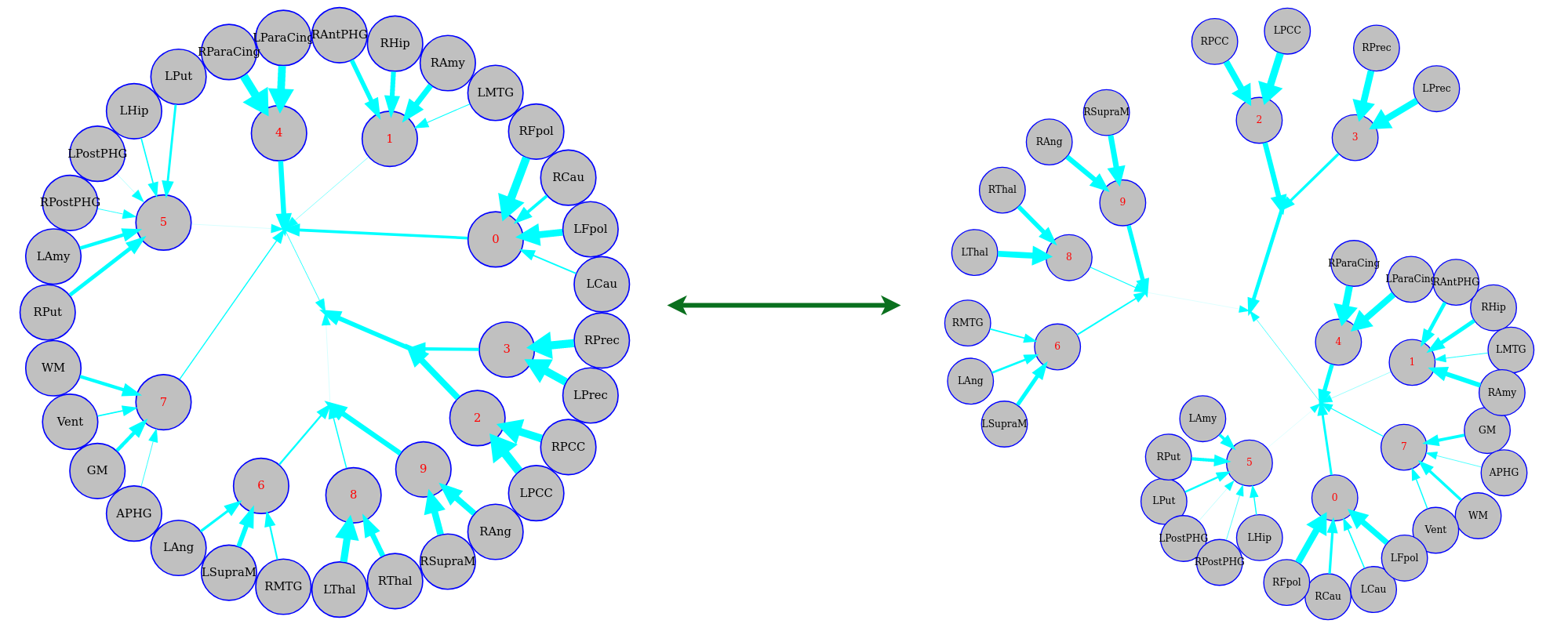}}}
    \caption{\textbf{Functional connectivity representation with graph-based networks.} Top row: the left and right figures correspond to Pearson correlation with a threshold of 0.14 and mutual information with a threshold of 0.02, respectively. Bottom row: the figures show total correlation with a threshold of 0.16 that was estimated by CorEx. To more directly display the statistical dependencies of brain regions, we here convert the circle graph to a tree graph. The weights are shown by the thickness of the edges, which shows how strongly information is coupled between or among brain regions.}
    \label{Fig:cc_mi_graph}
\end{figure}

Meanwhile, we also use weighted graph theory to cluster dependence among ROIs and we threshold edges with a weight of less than 0.16 for legibility with the CorEx approach. As we mentioned above, mutual information only estimates a more robust relationship between ROIs compared to correlation. However, when we go beyond pairwise ROIs, CorEx captures richer information among all ROIs (see Fig.~\ref{Fig:cc_mi_graph}(bottom row)). \blue{Here, we selected $m_{1} = 10$, $m_{2} = 3$, $m_{3} = 1$ as the latent dimensional for each layer in our estimate of TC with CorEx,} and their corresponding convergent curves are plotted in Fig.~\ref{Fig:tc_bound}, it shows total correlation lower bound stops increasing. Fig.~\ref{Fig:cc_mi_graph}(bottom row) shows the overall structure of the learned hierarchical model. Edge thickness is determined by $\alpha_{i, j} I\left(X_{i}: Y_{j}\right)$. The size of each node is proportional to the total correlation that a latent factor explains about its children. The discovered structure captures several significant relationships among ROIs that are consistent with correlation and mutual information results, e.g., LPCC and RPCC, LThal and RThal, LParaCing and RParaCing, LPut and RPut. Furthermore, TC discovered some beyond pairwise unknown relationships, for example, LCau, RCau, LFpol, and RFpol are clustered under node $0$, which explains why they have dense dependency during this cognitive task compared to other ROIs in the brain. 

\subsection{Large scale Connectome with resting-state fMRI}

\subsubsection{A selection of pre-defined atlas}
\begin{figure}[!ht]
    \centering
    \includegraphics[width=\textwidth, height=25em]{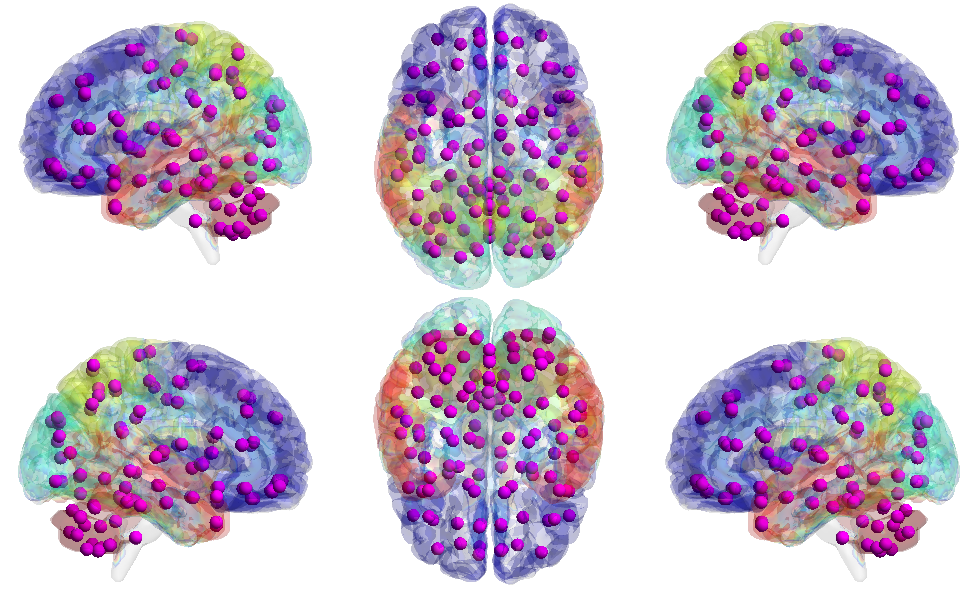}
    \caption{\textbf{Automated Anatomical Labeling (AAL) atlas.} The graph showed the volume of AAL (116 regions) mapped to the smoothed Colin27 brain surface template. The different brain areas are labeled on the brain surface with different colors, and detailed ROI/purple node information can be found in the Appendix section with Table~\ref{Tab:aal_rois_inf}.}
    \label{Fig:aal_rois}
\end{figure}

We use the Automated Anatomical Labeling (AAL) atlas~\cite{Tzourio02}, a structural atlas with 116 ROIs identified from the anatomy of a reference subject(see Fig.~\ref{Fig:aal_rois}.).

\subsubsection{Time series signals extraction}

HCP and ACPI can access raw and preprocessed data as well as phenotypic information about data samples. The raw rs-fMRI data was preprocessed using the Configurable Pipeline for the Analysis of Connectomes, an open-source software pipeline that allows for automated rs-fMRI data preprocessing and analysis. \blue{We extract time series for each ROI in each subject after defining anatomical brain ROIs with the AAL atlas}. We calculate the weighted average of the fMRI BOLD signals across all voxels in each region. Furthermore, the BOLD signal in each region is normalized and subsampled by repetition time. Finally, we average all of the subjects' time series signals in each ROI.

\subsubsection{HCP900}

\begin{figure}[htbp]
    \centering
    \subfloat[]{\scalebox{1}{\includegraphics[width=\textwidth, height=32em]{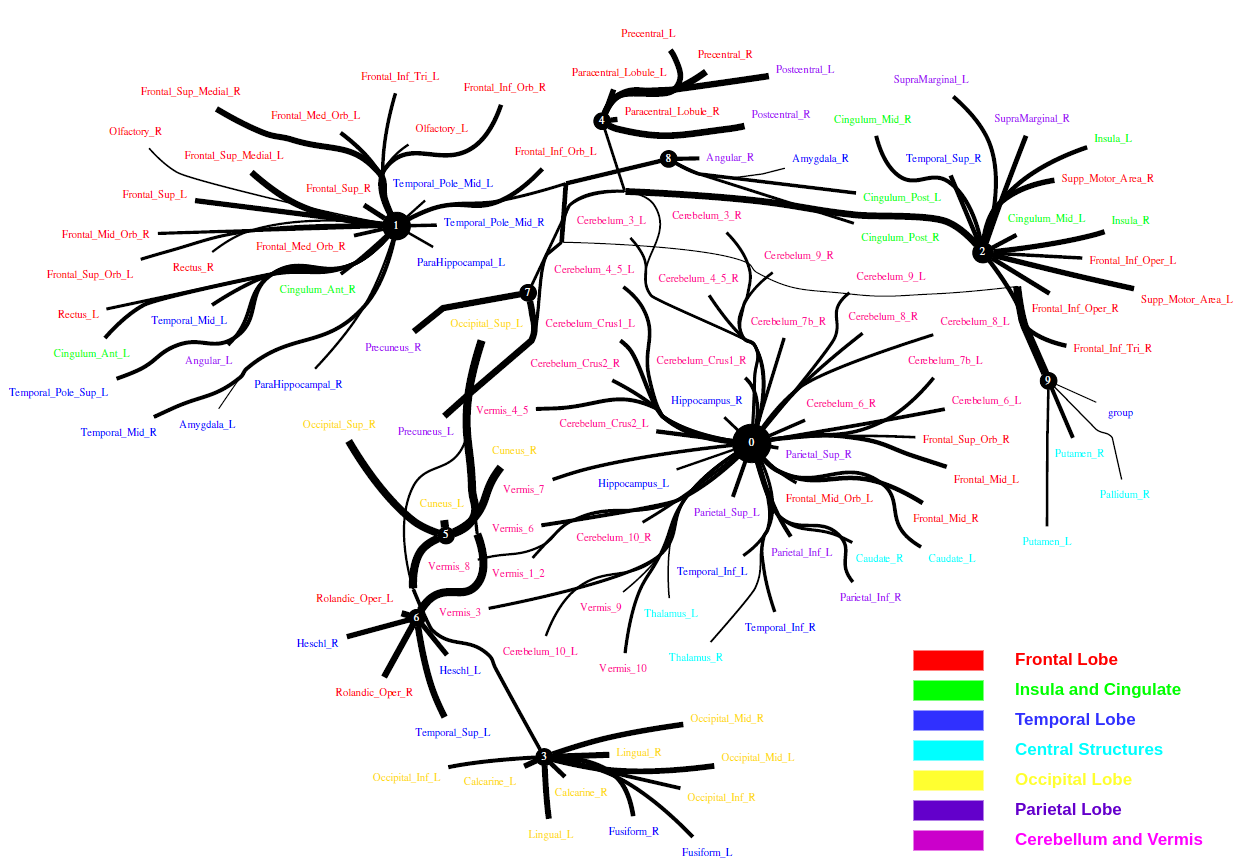}}} \\
    \subfloat[]{\scalebox{0.5}{\includegraphics[width=\textwidth, height=43em]{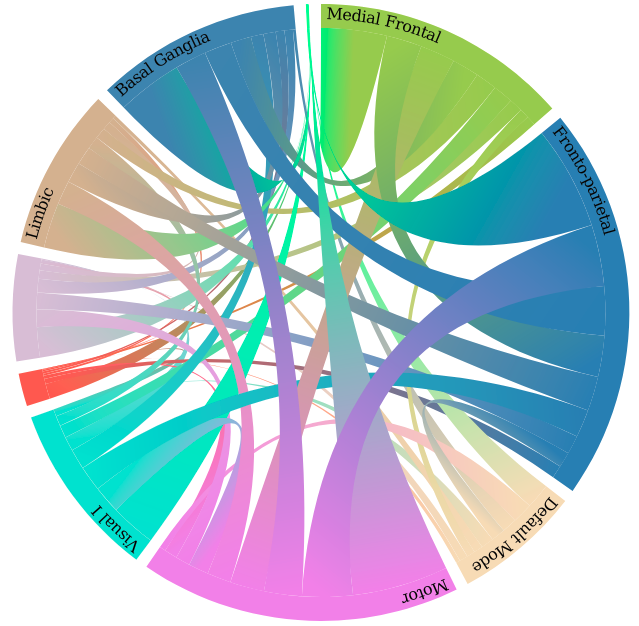}}} 
    \subfloat[]{\scalebox{0.5}{\includegraphics[width=\textwidth, height=43em]{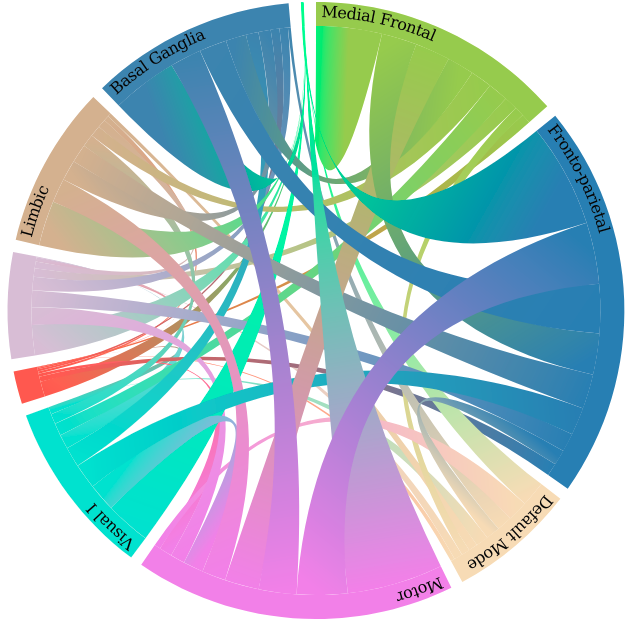}}} \\ 
    \caption{\textbf{Large scale functional connectivity with HCP900.} Top row: A weighted threshold graph with max of 86 edges showing the overall structure of the representation learned from AAL ROIs (\black{A high-resolution figure is represented in the appendix with Fig.11.}). Edge thickness is proportional to mutual information, and node size represents total correlation among children. In the node with red color, the frontal lobe is represented, while green color represents the insula and cingulate regions, blue color represents the temporal lobe, cyan color represents the central areas, gold color represents the occipital lobe, purple color represents the parietal lobe, and deep pink color represents the cerebellum and vermis. Bottom row: Two representative connectomes are presented in the form of a circular chord that shows the connections of all 116 nodes with (b) correlation and (c) mutual information of the HCP dataset. Each lobel was labeled with a different color.}
    \label{Fig:hcp_tc}
\end{figure}

The Human Connectome Project contains imaging and behavioral data from healthy people~\cite{Van13}. To investigate rest-state functional connectivity, we used preprocessed rest-fMRI data from the HCP900\footnote{https://www.humanconnectome.org/} release~\cite{Essen12}. \blue{Here, we selected $m_{1} = 10$, $m_{2} = 5$, $m_{3} = 1$ as the latent dimensional for each layer in our estimate of TC with CorEx.} We threshold edges with a weight of less than 0.16 for legibility. The Fig.~\ref{Fig:hcp_tc} has shown that whole brain resting-state functional connectivity is estimated with CorEx compared to Pearson correlation and mutual information. It mostly captures relationships among brain regions and neighboring brain regions cluster together and communicate with other areas, e.g., node $0$ has a bigger node size than other nodes. 

From Fig.~\ref{Fig:hcp_tc}, we found that brain regions are functionally clustered together, which is also consistent with structure connectivity based on their physical connectivity distance. For example, under node $0$, the cerebellum and vermis regions densely cluster together, while under node $1$, the frontal lobes cluster together and are also densely functionally connected with the temporal lobe, and so on. The different colors indicate different brain regions, which are based on Table.~\ref{Tab:aal_rois_inf}. In addition, we can see that functional integration and separation exist in our brain from Fig.~\ref{Fig:hcp_tc}. 

\subsubsection{Computational psychiatry applications with ACPI}

\begin{sidewaysfigure}
    \centering
    \subfloat[]{\scalebox{0.53}{\includegraphics[width=\textwidth,height=50em]{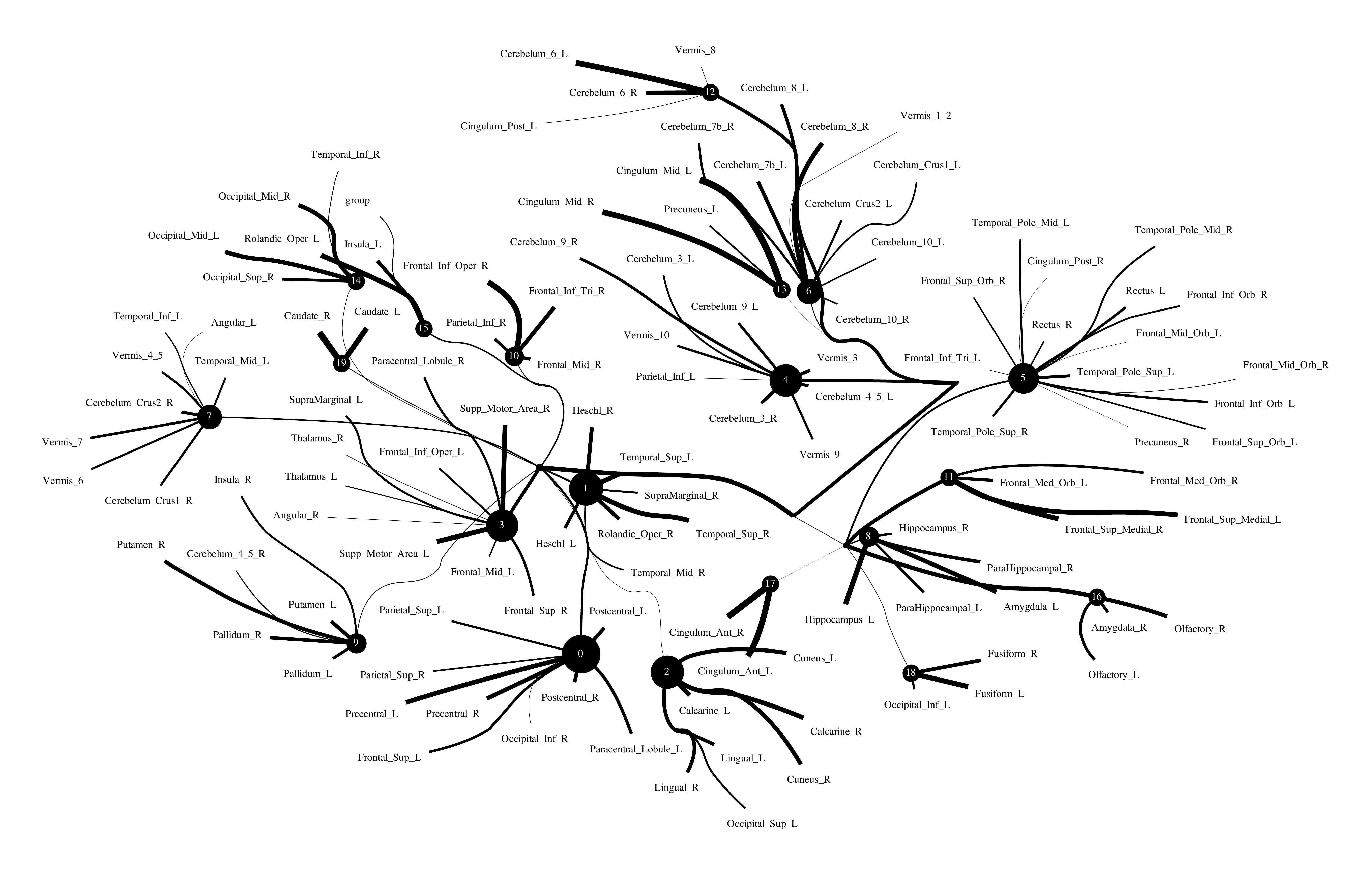}}} 
    \subfloat[]{\scalebox{0.53}{\includegraphics[width=\textwidth,height=50em]{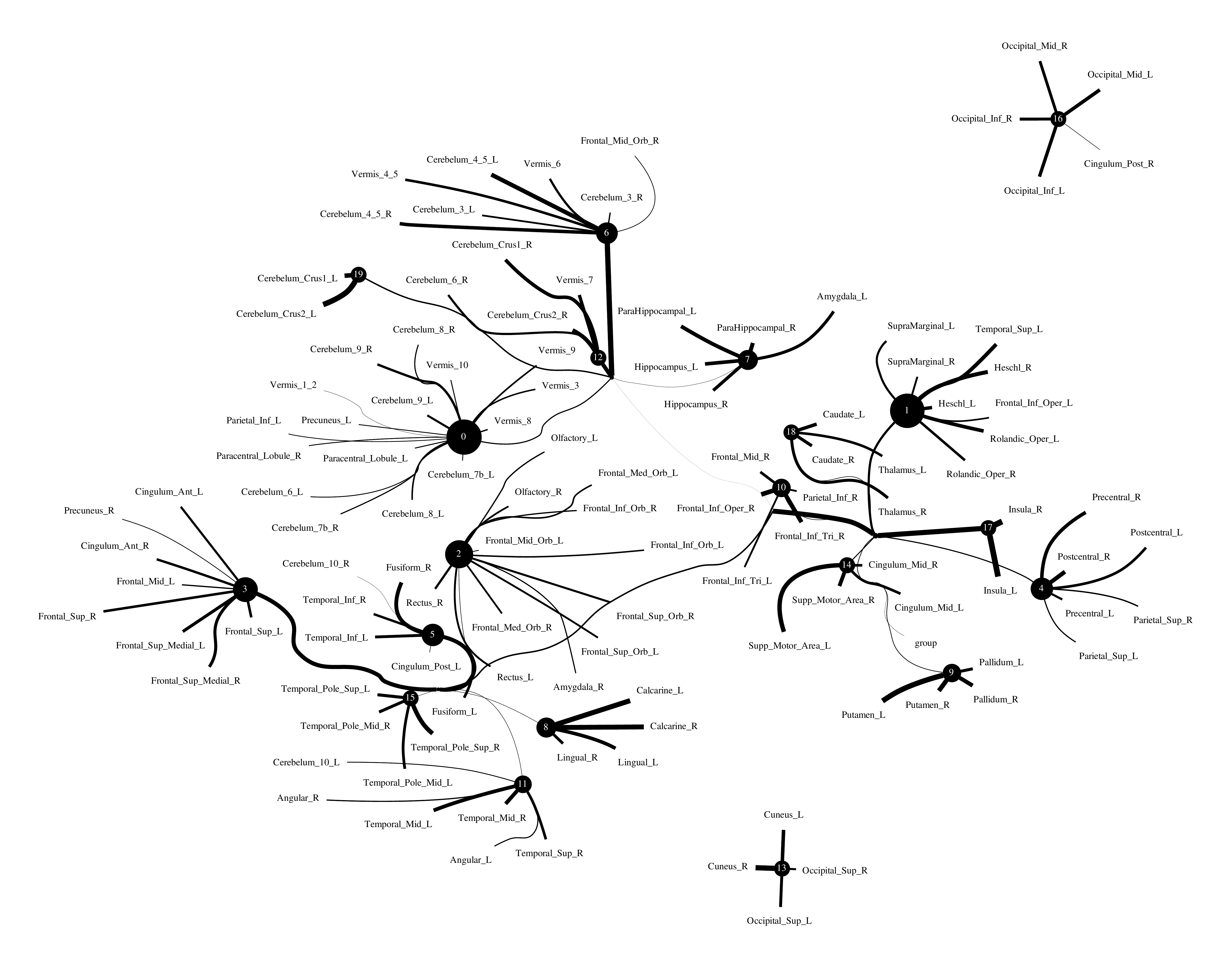}}}
    \caption{\textbf{Functional connectivity between health groups and patient groups.} A weighted threshold graph showing the overall structure of the representation learned from ALL ROIs. Edge thickness is proportional to mutual information, and node size represents total correlation among children. \blue{Here, we selected $m_{1} = 20$, $m_{2} = 3$, $m_{3} = 1$ as the latent dimensional for each layer in our estimate of TC with CorEx.} The figure (a) refers to normal people's functional connectivity and the figure (b) shows the marijuana group of functional connectivity in the brain. Both groups were measured with TC that used the same parameters in the model. In comparison with the healthy group, we found less functional connectivity happened in the patient group, e.g., Fronto-Parietal lobel and Default Model Regions. (\black{A high-resolution figure is represented in the appendix with Fig.12 and 13.})}
    \label{Fig:acpi_tc}
\end{sidewaysfigure}

The Addiction Connectome Preprocessed Initiative is a longitudinal study to investigate the effects of cannabis use among adults with a childhood diagnosis of ADHD. In particular, we use readily-preprocessed rest-fMRI data from the Multimodal Treatment Study of Attention Deficit Hyperactivity Disorder (MTA).\blue{We attempt to use functional connectivity as a bio-marker to discriminate whether individuals have consumed marijuana or not (62 marijuana group vs 64 control group). In a comparison of whole brain functional connectivity between control and patient groups, we found altered functional connectivity in the patient group compared to the healthy group (see Fig.~\ref{Fig:acpi_tc}.). We quantify the difference between patient groups compared to healthy groups, and the purity of patient groups compared to control groups is $0.85 \pm 0.23$. The significant altered functional connectivity happened between the frontoparietal and motor regions. Meanwhile, we found sparse functional connectivity in the patient group compared to the control group in general. Meanwhile, we also discovered that marijuana users had more interaction between neural time series in particular ROIs such as the cerebellum, fronto-parietal, and default model regions than controls, e.g., cerebellum regions mainly densely cluster around node$0$ compared to the control group.} 
It also may explain differences in behavior in marijuana users 
because the fronto-parietal network controls cognitive behavior execution and decision-making, cerebellum-related action, and default model network dysfunction in addiction users. All the above results are consistent with previous related research \cite{Behan13, Bubl07,Rui19}. Moreover, we found some unknown disconnect between some visual regions and other brain areas. Based on related similarity research \cite{Giedd08, Medina07}, we suggest that marijuana patients may alter visual perception too. 

\section{Discussion}

This manuscript presents a higher-order information-theoretic measure to estimate functional connectivity. We estimated total correlation with CorEx under different situations. However, the approach has its own pros and cons, which we will discuss later. Furthermore, we found total correlation can be a metric to estimate functional connectivity in the human brain. It can identify some well-known functional connectivities and capture a few unknown nonlinear relationships among brain regions as well. To the best of our knowledge, this is the first time that total correlation has been used to estimate larger-scale functional connectivity for a whole-brain-AAL atlas with 116 structural ROIs. Total correlation can also be a tool to find biomarkers to help us diagnose brain-related diseases.

Here, we will discuss some advantages and limitations of this research now. Firstly, given the curse of dimensionality of fMRI, we need to find a low-dimensional representation that helps us characterize the connectivity. Traditional general linear models (GLM), such as expert-defined ROIs or the ALL atlas, are frequently used to find ROIs in resting-state experiments. However, we should be able to do better with a data-driven approach. Sample sizes and statistical thresholds are known to have a major impact on the statistical power and accuracy of GLM-based ROI selection. Previous research has revealed that GLM has limited statistical power when inferring from fMRI data~\cite{Poline12,DOWDLE2021102171}. However, we used GLM-based ROI selection in the real fMRI datasets, which may affect the final result when we estimate functional connectivity.

Second, CorEx is model-independent, which means no anatomical or functional prior knowledge is required to estimate ROIs. The method is entirely data-driven; this way, it is possible to analyze networks that have not been investigated and could be a future extension of work. It is also possible to use total correlation as a pre-analysis for other techniques like dynamic causal modeling, which need constraints about the underlying network~\cite{Marreiros10}. 
What differentiates the CorEx algorithm is that it tries to break the variables into clusters with high TC. 
In other words, \black{CoRex finds a tree of latent factors that explain the total correlation, 
so this tree of clusters based on TC is a more data-driven way to define regions and then connectivity
than ROIs predefined by hand.} This prioritization of ''modular'' solutions in CorEx was not realized or emphasized in the original research. The second reason why we used CorEx to estimate functional connectivity on larger-scale fMRI datasets is that it is a clustering approach via TC. Furthermore, CorEx estimates total correlation via hierarchical maximization correlation between previous layer and current layer variables with a tight information bound that push mode estimates a more accurate relationship among variables in real neural signals. 

\black{Third}, TC is an indirect information quantitative tool that cannot determine the direction of information flow between brain regions. Meanwhile, we discovered some unknown functional connectivity in the real fMRI dataset before. 

\black{Fourth,} \blue{given the irregularity of neural time series and the difficulties in quantifying graph signals when brain networks are represented by graphs, we should avoid quantifying too many graph signals. However, there is a metric called permutation entropy that gives us the possibility to quantify the graph signal in complex systems~\cite{Fabila22}. It could be very interesting to apply this metric to brain networks to check how much information could be obtained from the complex graph signals, which could then help us more deeply understand brain networks in the future. Moreover, as we mentioned the complexity of neural time series, one of the important potential problems is the length of time series, except for the additional dimensional problem. It's a significant challenge when you are processing long lengths of time series, but it could be solved by transforming the time series into embedding space or segmenting the long time series into specific time windows~\cite{Porta16}.}

Finally, we applied TC to estimate large scale functional connectivity with the real fMRI dataset across HCP and ACPI. The functional connectivity with HCP900 gives us the potential to estimate a full brain atlas with TC in the future, and our result shows that TC can capture the right functional connectivity, and beyond this, it could also give us some unknown functional connectivity. Therefore, it could be a future extension project. Furthermore, we used TC as a possible method to find biomarkers of brain disease with the ACPI dataset. We compared whole-brain functional connectivity between control and patient groups. We found altered functional connectivity in the patient group compared to the healthy group, and we quantified this difference with purity metrics because it’s a simple and transparent evaluation measure. The purity in patient groups compared to control groups is not too large, and it shows that there is some altered functional connectivity in the patient group, for instance, we mentioned brain networks in the cerebellum, fronto-parietal, and default model regions. However, it's just examined with one dataset with \black{small number of subjects} and does not consider within-subjects variability, and could be extended with more large datasets in the future.

\section{Conclusions}

We have introduced total correlation to capture multivariate large scale interactions within brain regions. 
They have been experimentally verified as effective steps for reconstructing multivariate relationships in the brain. 
In this study, CorEx was adopted to estimate total correlation. The CorEx approach can capture functional connectivity characteristics when going beyond pairwise brain regions. \blue{On the other hand, we evaluated the method with resting-state fMRI datasets.} We found that multivariable relationships cannot be detected if we use pairwise correlation and mutual information quantities only. More generally, multivariable relationships can be clustered only if we use total correlation. Therefore, total correlation measures are significant to find complicated functional connectivity among brain regions. Also, we have shown that total correlation can estimate functional connectivity in the real neural dataset and find biomarkers for diagnosing brain diseases.

\textcolor{black}{In the future, we plan to use the functional connectivity relationships discovered by total correlation as an input to existing graph neural networks (GNNs)~\cite{welling2016semi} for the purpose of interpretable brain disease diagnosis, such that practitioners or doctors can identify the most informative subgraphs (or modules) to the decision (e.g., autism patients or health-control groups). 
\black{In this regard, quantitative measures to define differences between graphs~\cite{Tantardini19}, and extension of analytical results in~\cite{Li22a} to larger number of nodes will be critical to assess and improve the qualitative results presented here.}
The recently proposed approaches (e.g., \cite{cui2021brainnnexplainer,zheng2022brainib}) all rely on pairwise relationships estimated by linear correlation coefficient as the input, which ignores high-order dependence essentially. In this sense, we believe our approach has the potential to improve the explanation performances of existing GNN explainers on brains.}

\section*{Data Availability}

\blue{Data and code needed to reproduce the results presented here are available at \url{https://forms.gle/1DXDpEpi7AodQ77q7}.}

\section*{Acknowledgement}

QL and JM \black{were} partially funded by these spanish/european grants from GVA/AEI/FEDER/EU: MICINN PID2020-118071GB-I00, MICINN PDC2021-121522-C21, and GVA Grisolía-P/2019/035. GVS acknowledges support from the Defense Advanced Research Projects Agency (DARPA) under award FA8750-17-C-0106. SY was funded by the Research Council of Norway under grant no.309439. Final, we thank the organizers of the HCP, and ACPI for providing these interesting dataset which used in this studies.

\section*{Author Contributions:} Conceptualization, methodology, software and validation, \black{QL}. Writing—original draft preparation, writing—review, and editing, QL., GVS., SY., and JM. \black{Contribution of JM was focused on the definition of paper scope about large scale connectome, the relation with analytical results in~\cite{Li22a}, and the criticism of performance measures.}
All authors have read and agreed to the published version of the manuscript.

\section*{Conflicts of Interest:} The authors declare no conflict of interest. 

\section*{Abbreviations}
\blue{TC: Total Correlation\\ 
CorEx: Correlation Explanation\\
CC:Linear Correlation\\
I:Mutual Information\\
VAEs: Variational Autoencoders \\ 
fMRI: functional Magnetic Resonance Imaging\\
BOLD: Blood-Oxygen-Level-Dependent Imaging\\
DCM: Dynamic Causal Modeling \\
GLM: General Linear Model\\
ROI: Region of Interest\\
HCP:Human Connectome Project\\
MTA: Multimodal Treatment of Attention Deficit Hyperactivity Disorder\\
GNNs: Graph Neural Networks\\
}
\bibliographystyle{unsrt}  
\bibliography{main}

\begin{thebibliography}{10}

\bibitem{Friston11}
Karl Friston.
\newblock Functional and effective connectivity: A review.
\newblock {\em Brain connectivity}, 1:13--36, 01 2011.

\bibitem{Porta14}
Alberto Porta, Luca Faes, Vlasta Bari, Andrea Marchi, Tito Bassani,
  Giandomenico Nollo, Natália~Maria Perseguini, Juliana Milan, Vinícius
  Minatel, Audrey Borghi-Silva, Anielle C.~M. Takahashi, and Aparecida~M.
  Catai.
\newblock Effect of age on complexity and causality of the cardiovascular
  control: Comparison between model-based and model-free approaches.
\newblock {\em PLOS ONE}, 9(2):1--14, 02 2014.

\bibitem{Heuvel10}
Martijn Heuvel and Hilleke Pol.
\newblock Exploring the brain network: A review on resting-state fmri
  functional connectivity.
\newblock {\em European neuropsychopharmacology : the journal of the European
  College of Neuropsychopharmacology}, 20:519--34, 08 2010.

\bibitem{Sporns05}
Olaf Sporns, Giulio Tononi, and Rolf Kötter.
\newblock The human connectome: A structural description of the human brain.
\newblock {\em PLoS computational biology}, 1:e42, 10 2005.

\bibitem{Bastos16}
Andre Bastos and Jan-Mathijs Schoffelen.
\newblock A tutorial review of functional connectivity analysis methods and
  their interpretational pitfalls.
\newblock {\em Frontiers in Systems Neuroscience}, 9, 01 2016.

\bibitem{Lizier11}
J.T. Lizier, J.~Heinzle, A.~Horstmann, J.~Haynes, and M.~Prokopenko.
\newblock Multivariate information-theoretic measures reveal directed
  information structure and task relevant changes in f{MRI} connectivity.
\newblock {\em J. Comput. Neurosci.}, 30(1):85–107, 2011.

\bibitem{Piasini19}
Eugenio Piasini and Stefano Panzeri.
\newblock Information theory in neuroscience.
\newblock {\em Entropy}, 21:62, 01 2019.

\bibitem{Ince16}
Robin Ince, Bruno Giordano, Christoph Kayser, Guillaume Rousselet, Joachim
  Gross, and Philippe Schyns.
\newblock A statistical framework for neuroimaging data analysis based on
  mutual information estimated via a gaussian copula.
\newblock {\em Human brain mapping}, 38, 11 2016.

\bibitem{Dimitrov11}
Alexander Dimitrov, Aurel Lazar, and Jonathan Victor.
\newblock Information theory in neuroscience.
\newblock {\em Journal of computational neuroscience}, 30:1--5, 02 2011.

\bibitem{Borst99}
Alexander Borst and Frédéric Theunissen.
\newblock Information theory and neural coding.
\newblock {\em Nature neuroscience}, 2:947--57, 12 1999.

\bibitem{Bialek12}
Gasper Tkacik, Olivier Marre, Thierry Mora, Dario Amodei, Michael Berry~II, and
  William Bialek.
\newblock The simplest maximum entropy model for collective behavior in a
  neural network.
\newblock {\em Journal of Statistical Mechanics: Theory and Experiment}, 2013,
  07 2012.

\bibitem{Gomez19}
A.~Gomez-Villa, M.~Bertalmio, and J.~Malo.
\newblock Visual information flow in {W}ilson-{C}owan networks.
\newblock {\em J. Neurophysiol. doi: 10.1152/jn.00487.2019}, 2020.

\bibitem{Malo20}
J.~Malo.
\newblock Spatio-chromatic information available from different neural layers
  via gaussianization.
\newblock {\em J. Math. Neurosci.}, 10(18), 2020.

\bibitem{Malo22}
Jesús Malo.
\newblock Information flow in biological networks for color vision.
\newblock {\em Entropy}, 24(10), 2022.

\bibitem{Farahani19}
Farzad Farahani, Waldemar Karwowski, and Nichole Lighthall.
\newblock Application of graph theory for identifying connectivity patterns in
  human brain networks: A systematic review.
\newblock {\em Frontiers in Neuroscience}, 13:585, 06 2019.

\bibitem{Sporns18}
Olaf Sporns.
\newblock Graph theory methods: applications in brain networks.
\newblock {\em Dialogues in Clinical Neuroscience}, 20:111 -- 121, 2018.

\bibitem{Rosas18}
Fernando Rosas, Pedro~A.M. Mediano, Martín Ugarte, and Henrik~J. Jensen.
\newblock An information-theoretic approach to self-organisation: Emergence of
  complex interdependencies in coupled dynamical systems.
\newblock {\em Entropy}, 20(10), 2018.

\bibitem{Rosas19}
Fernando~E. Rosas, Pedro A.~M. Mediano, Michael Gastpar, and Henrik~J. Jensen.
\newblock Quantifying high-order interdependencies via multivariate extensions
  of the mutual information.
\newblock {\em Phys. Rev. E}, 100:032305, Sep 2019.

\bibitem{Tononi99}
Giulio Tononi and Gerald Edelman.
\newblock Consciousness and complexity.
\newblock {\em Science (New York, N.Y.)}, 282:1846--51, 01 1999.

\bibitem{Pereda05}
Ernesto Pereda, Rodrigo Quian, and Joydeep Bhattacharya.
\newblock Nonlinear multivariate analysis of neurophysiological signals.
\newblock {\em Progress in neurobiology}, 77:1--37, 09 2005.

\bibitem{Chai09}
Barry Chai, Dirk~B. Walther, Diane~M. Beck, and Li~Fei-Fei.
\newblock Exploring functional connectivity of the human brain using
  multivariate information analysis.
\newblock In {\em Proceedings of the 22nd International Conference on Neural
  Information Processing Systems}, NIPS'09, page 270–278, Red Hook, NY, USA,
  2009. Curran Associates Inc.

\bibitem{Wang15}
Zhe Wang, Ahmed Alahmadi, David Zhu, and Tongtong li.
\newblock Brain functional connectivity analysis using mutual information.
\newblock In {\em 2015 IEEE Global Conference on Signal and Information
  Processing (GlobalSIP)}, pages 542--546, 12 2015.

\bibitem{Jomaa19}
Mohamad El~Sayed Hussein~Jomaa, Marcelo Colominas, Nisrine Jrad, Patrick
  Van~Bogaert, and Anne Humeau-Heurtier.
\newblock A new mutual information measure to estimate functional connectivity:
  Preliminary study.
\newblock In {\em Conference proceedings: Annual International Conference of
  the IEEE Engineering in Medicine and Biology Society}, volume 2019, pages
  640--643, 07 2019.

\bibitem{Li22b}
Qiang Li.
\newblock Functional connectivity inference from fmri data using multivariate
  information measures.
\newblock {\em Neural Networks}, 146:85--97, 2022.

\bibitem{Li22a}
Qiang Li, Greg Ver~Steeg, and Jesus Malo.
\newblock Functional connectivity in visual areas from total correlation.
\newblock {\em ArXiV https://arxiv.org/abs/2208.05770}, 08 2022.

\bibitem{Watanabe60}
Satosi Watanabe.
\newblock Information theoretical analysis of multivariate correlation.
\newblock {\em IBM Journal of research and development}, 4(1):66--82, 1960.

\bibitem{Studeny98}
M.~Studeny and J.~Vejnarova.
\newblock The multi-information function as a tool for measuring stochastic
  dependence.
\newblock {\em Learning in graphical models}, pages 261--298, January 1998.

\bibitem{Laparra11}
V.~Laparra, G.~Camps-Valls, and J.~Malo.
\newblock Iterative gaussianization: from {ICA} to random rotations.
\newblock {\em IEEE Trans. Neural Networks}, 22(4):537--549, 2011.

\bibitem{laparra2020}
V.~Laparra, E.~Johnson, G.~Camps, R.~Santos, and J.~Malo.
\newblock Information theory measures via multidimensional gaussianization.
\newblock {\em ArXiV: Stats. Mach. Learn.}, page
  https://arxiv.org/abs/2010.03807, 2020.

\bibitem{Van13}
David Van~Essen, Stephen Smith, Deanna Barch, Timothy Behrens, Essa Yacoub, and
  Kamil Ugurbil.
\newblock The wu-minn human connectome project: an overview.
\newblock {\em NeuroImage}, 80, 05 2013.

\bibitem{Essen12}
D.C. Essen, K~Ugurbil, Edward Auerbach, Deanna Barch, T.E.J. Behrens, Richard
  Bucholz, A~Chang, Liyong Chen, Maurizio Corbetta, Sandra Curtiss, Stefania
  Della~Penna, David Feinberg, Matthew Glasser, Noam Harel, A.C. Heath, Linda
  Larson-Prior, Daniel Marcus, Georgios Michalareas, Steen Moeller, and Essa
  Yacoub.
\newblock The human connectome project: A data acquisition perspective.
\newblock {\em NeuroImage}, 62:2222--31, 02 2012.

\bibitem{steeg2014NIPS}
Greg~Ver Steeg and Aram Galstyan.
\newblock Discovering structure in high-dimensional data through correlation
  explanation.
\newblock In {\em Advances in Neural Information Processing Systems,
  NIPS{\textquoteright}14}, 2014.

\bibitem{steeg2015corex_theory}
Greg~Ver Steeg and Aram Galstyan.
\newblock Maximally informative hierarchical representations of
  high-dimensional data.
\newblock In {\em AISTATS{\textquoteright}15}, 2015.

\bibitem{Cover06}
Thomas~M. Cover and Joy~A. Thomas.
\newblock {\em Elements of Information Theory (Wiley Series in
  Telecommunications and Signal Processing)}.
\newblock Wiley-Interscience, USA, 2006.

\bibitem{Kraskov04}
Alexander Kraskov, Harald St\"ogbauer, and Peter Grassberger.
\newblock Estimating mutual information.
\newblock {\em Phys. Rev. E}, 69:066138, Jun 2004.

\bibitem{Fabila22}
John~Stewart Fabila-Carrasco, Chao Tan, and Javier Escudero.
\newblock Permutation entropy for graph signals.
\newblock {\em IEEE Transactions on Signal and Information Processing over
  Networks}, 8:288--300, 2022.

\bibitem{Lyu09}
Siwei Lyu and Eero~P. Simoncelli.
\newblock {Nonlinear Extraction of Independent Components of Natural Images
  Using Radial Gaussianization}.
\newblock {\em Neural Computation}, 21(6):1485--1519, 2009.

\bibitem{aistats2019}
Shuyang Gao, Robert Brekelmans, Greg {Ver Steeg}, and Aram Galstyan.
\newblock Auto-encoding correlation explanation.
\newblock In {\em Proceedings of the 22nd International Conference on AI and
  Statistics (AISTATS)}, 2019.

\bibitem{yu2019multivariate}
Shujian Yu, Luis Gonzalo~Sanchez Giraldo, Robert Jenssen, and Jose~C Principe.
\newblock Multivariate extension of matrix-based r{\'e}nyi's $\alpha$-order
  entropy functional.
\newblock {\em IEEE transactions on pattern analysis and machine intelligence},
  42(11):2960--2966, 2019.

\bibitem{Shannon48}
Claude~E. Shannon.
\newblock A mathematical theory of communication.
\newblock {\em Bell Syst. Tech. J.}, 27(3):379--423, 1948.

\bibitem{steeg2017unsupervised}
Greg Ver~Steeg.
\newblock Unsupervised learning via total correlation explanation.
\newblock In {\em IJCAI}, 2017.

\bibitem{ver2019fast}
Greg Ver~Steeg, Hrayr Harutyunyan, Daniel Moyer, and Aram Galstyan.
\newblock Fast structure learning with modular regularization.
\newblock In {\em Advances in Neural Information Processing Systems}, pages
  15567--15577, 2019.

\bibitem{manning08}
Christopher~D. Manning, Prabhakar Raghavan, and Hinrich Schütze.
\newblock {\em Introduction to Information Retrieval}.
\newblock Cambridge University Press, Cambridge, UK, 2008.

\bibitem{Tzourio02}
Nathalie Tzourio-Mazoyer, Brigitte Landeau, Papathanassiou DF, Fabrice
  Crivello, O.N.D. Etard, Nicolas Delcroix, Bernard Mazoyer, and Joliot Marc.
\newblock Automated anatomical labeling of activations in spm using a
  macroscopic anatomical parcellation of the mni mri single-subject brain.
\newblock {\em NeuroImage}, 15:273--89, 02 2002.

\bibitem{Behan13}
B~Behan, Gregory Connolly, S~Datwani, M~Doucet, J~Ivanovic, R~Morioka, A~Stone,
  Richard Watts, B~Smyth, and Hugh Garavan.
\newblock Response inhibition and elevated parietal-cerebellar correlations in
  chronic adolescent cannabis users.
\newblock {\em Neuropharmacology}, 84, 06 2013.

\bibitem{Bubl07}
Emanuel Bubl, Ludger Tebartz~van Elst, Matthias Gondan, Dieter Ebert, and Mark
  Greenlee.
\newblock Vision in depressive disorder.
\newblock {\em The world journal of biological psychiatry : the official
  journal of the World Federation of Societies of Biological Psychiatry},
  10:377--84, 10 2007.

\bibitem{Rui19}
Rui Zhang and Nora Volkow.
\newblock Brain default-mode network dysfunction in addiction.
\newblock {\em NeuroImage}, 200, 06 2019.

\bibitem{Giedd08}
Jay Giedd, Matcheri Keshavan, and Tomas Paus.
\newblock Why do many psychiatric disorders emerge during adolescence?
\newblock {\em Nat Rev Neurosci}, 9:947--957, 01 2008.

\bibitem{Medina07}
Krista Medina, Karen Hanson, Alecia Dager, Mairav Cohen-Zion, Bonnie Nagel, and
  Susan Tapert.
\newblock Neuropsychological functioning in adolescent marijuana users: Subtle
  deficits detectable after a month of abstinence.
\newblock {\em Journal of the International Neuropsychological Society : JINS},
  13:807--20, 09 2007.

\bibitem{Poline12}
Jean-Baptiste Poline and Matthew Brett.
\newblock The general linear model and fmri: Does love last forever?
\newblock {\em NeuroImage}, 62:871--80, 02 2012.

\bibitem{DOWDLE2021102171}
Logan~T. Dowdle, Geoffrey Ghose, Clark~C.C. Chen, Kamil Ugurbil, Essa Yacoub,
  and Luca Vizioli.
\newblock Statistical power or more precise insights into neuro-temporal
  dynamics? assessing the benefits of rapid temporal sampling in fmri.
\newblock {\em Progress in Neurobiology}, 207:102171, 2021.

\bibitem{Marreiros10}
André Marreiros, Klaas Stephan, and Karl Friston.
\newblock Dynamic causal modeling.
\newblock {\em Scholarpedia}, 5:9568, 07 2010.

\bibitem{Porta16}
Alberto Porta and Luca Faes.
\newblock Wiener–granger causality in network physiology with applications to
  cardiovascular control and neuroscience.
\newblock {\em Proceedings of the IEEE}, 104:282--309, 2016.

\bibitem{welling2016semi}
Max Welling and Thomas~N Kipf.
\newblock Semi-supervised classification with graph convolutional networks.
\newblock In {\em J. International Conference on Learning Representations (ICLR
  2017)}, 2016.

\bibitem{Tantardini19}
M.~Tantardini, F.~Ieva, L.~Tajoli, and C.~Piccardi.
\newblock Comparing methods for comparing networks.
\newblock {\em Sci. Rep.}, 9:17557, 2019.

\bibitem{cui2021brainnnexplainer}
Hejie Cui, Wei Dai, Yanqiao Zhu, Xiaoxiao Li, Lifang He, and Carl Yang.
\newblock Brainnnexplainer: An interpretable graph neural network framework for
  brain network based disease analysis.
\newblock {\em arXiv preprint arXiv:2107.05097}, 2021.

\bibitem{zheng2022brainib}
Kaizhong Zheng, Shujian Yu, Baojuan Li, Robert Jenssen, and Badong Chen.
\newblock Brainib: Interpretable brain network-based psychiatric diagnosis with
  graph information bottleneck.
\newblock {\em arXiv preprint arXiv:2205.03612}, 2022.

\end{thebibliography}

\newpage
\section{Appendix}
\begin{table}[H]
\centering
\caption{Information of 116 brain regions that are comprised in the AAL atlas}
\resizebox{0.72\textwidth}{!}{%
\begin{tabular}{lll}
\hline \hline
Brain Area   & AAL Regions                                           & AAL Index No. \\ \hline \hline
             & Precentral gyrus                                      & 1, 2           \\ 
             & Superior frontal gyrus, dorsolateral                  & 3, 4           \\ 
             & Superior frontal gyrus, orbital part                  & 5, 6           \\ 
             & Middle frontal gyrus                                  & 7, 8           \\ 
             & Middle frontal gyrus, orbital part                    & 9, 10          \\ 
             & Inferior frontal gyrus, opercular part                & 11, 12         \\ 
             & Inferior frontal gyrus, triangular part               & 13, 14         \\ 
Frontal Lobe & Inferior frontal gyrus, orbital part                  & 15, 16         \\ 
             & Rolandic operculum                                    & 17, 18         \\ 
             & Supplementary motor area                              & 19, 20         \\ 
             & Olfactory cortex                                      & 21, 22         \\ 
             & Superior frontal gyrus, medial                        & 23, 24         \\ 
             & Superior frontal gyrus, medial orbital                & 25, 26         \\ 
             & Gyrus rectus                                          & 27, 28         \\ 
             & Paracentral lobule                                    & 69, 70         \\ \hline \hline
             & Insula                                                & 29, 30         \\ 
Insula and   & Anterior cingulate and paracingulate gyri             & 31, 32         \\ 
Cingulate    & Median cingulate and paracingulate gyri               & 33, 34         \\ 
             & Posterior cingulate gyrus                             & 35, 36         \\ \hline \hline
             & Hippocampus                                           & 37, 38         \\ 
             & Parahippocampal gyrus                                 & 39, 40         \\ 
             & Amygdala                                              & 41, 42         \\ 
             & Fusiform gyrus                                        & 55, 56         \\ 
Temporal     & Heschl gyrus                                          & 79, 80         \\ 
Lobe         & Superior temporal gyrus                               & 81, 82         \\ 
             & Temporal pole: superior temporal gyrus                & 83, 84         \\ 
             & Middle temporal gyrus                                 & 85, 86         \\ 
             & Temporal pole: middle temporal gyrus                  & 87, 88         \\ 
             & Inferior temporal gyrus                               & 89, 90         \\ \hline \hline
             & Caudate nucleus                                       & 71, 72         \\ 
Central      & Lenticular nucleus, putamen                           & 73, 74         \\ 
Structures   & Lenticular nucleus, pallidum                          & 75, 76         \\ 
             & Thalamus                                              & 77, 78         \\ \hline \hline 
             & Calcarine fissure and surrounding cortex              & 43, 44         \\ 
             & Cuneus                                                & 45, 46         \\ 
Occipital    & Lingual gyrus                                         & 47, 48         \\ 
Lobe         & Superior occipital gyrus                              & 49, 50         \\ 
             & Middle occipital gyrus                                & 51, 52         \\ 
             & Inferior occipital gyrus                              & 53, 54         \\ \hline \hline
             & Postcentral gyrus                                     & 57, 58         \\ 
             & Superior parietal gyrus                               & 59, 60         \\ 
Parietal     & Inferior parietal, but supramarginal and angular gyri & 61, 62         \\ 
Lobe         & Supramarginal gyrus                                   & 63, 64         \\ 
             & Angular gyrus                                         & 65, 66         \\ 
             & Precuneus                                             & 67, 68         \\ \hline \hline 
             & Cerebelum Crus 1                                      & 91, 92         \\ 
             & Cerebelum Crus 2                                      & 93, 94         \\ 
             & Cerebelum 3                                           & 95, 96         \\ 
             & Cerebelum 4, 5                                        & 97, 98         \\ 
             & Cerebelum 6                                           & 99, 100        \\ 
             & Cerebelum 7b                                          & 101, 102       \\ 
             & Cerebelum 8                                           & 103, 104       \\ 
             & Cerebelum 9                                           & 105, 106       \\ 
Cerebellum 
and Vermis  
             & Cerebelum 10                                          & 107, 108       \\ 
             & Vermis 1, 2                                           & 109            \\ 
             & Vermis 3                                              & 110            \\ 
             & Vermis 4, 5                                           & 111            \\ 
             & Vermis 6                                              & 112            \\ 
             & Vermis 7                                              & 113            \\ 
             & Vermis 8                                              & 114            \\ 
             & Vermis 9                                              & 115            \\ 
             & Vermis 10                                             & 116            \\ \hline \hline
\end{tabular}
}
\label{Tab:aal_rois_inf}
\end{table}

\newpage
\begin{sidewaysfigure}
    \centering
    {\scalebox{1}{\includegraphics[width=\textwidth,height=50em]{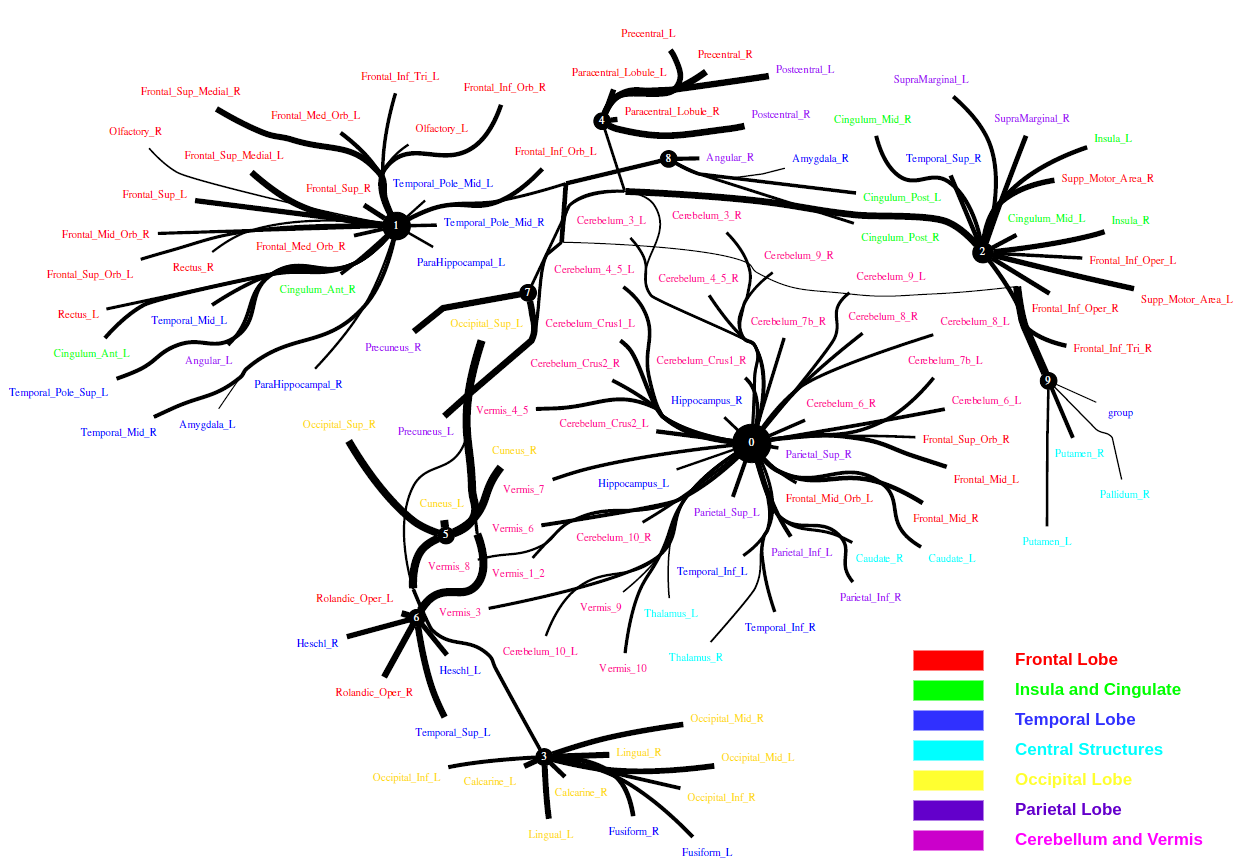}}}
    \caption{\textbf{Functional connectivity of HCP900.}} 
\end{sidewaysfigure}

\newpage
\begin{sidewaysfigure}
    \centering
    {\scalebox{1}{\includegraphics[width=\textwidth,height=50em]{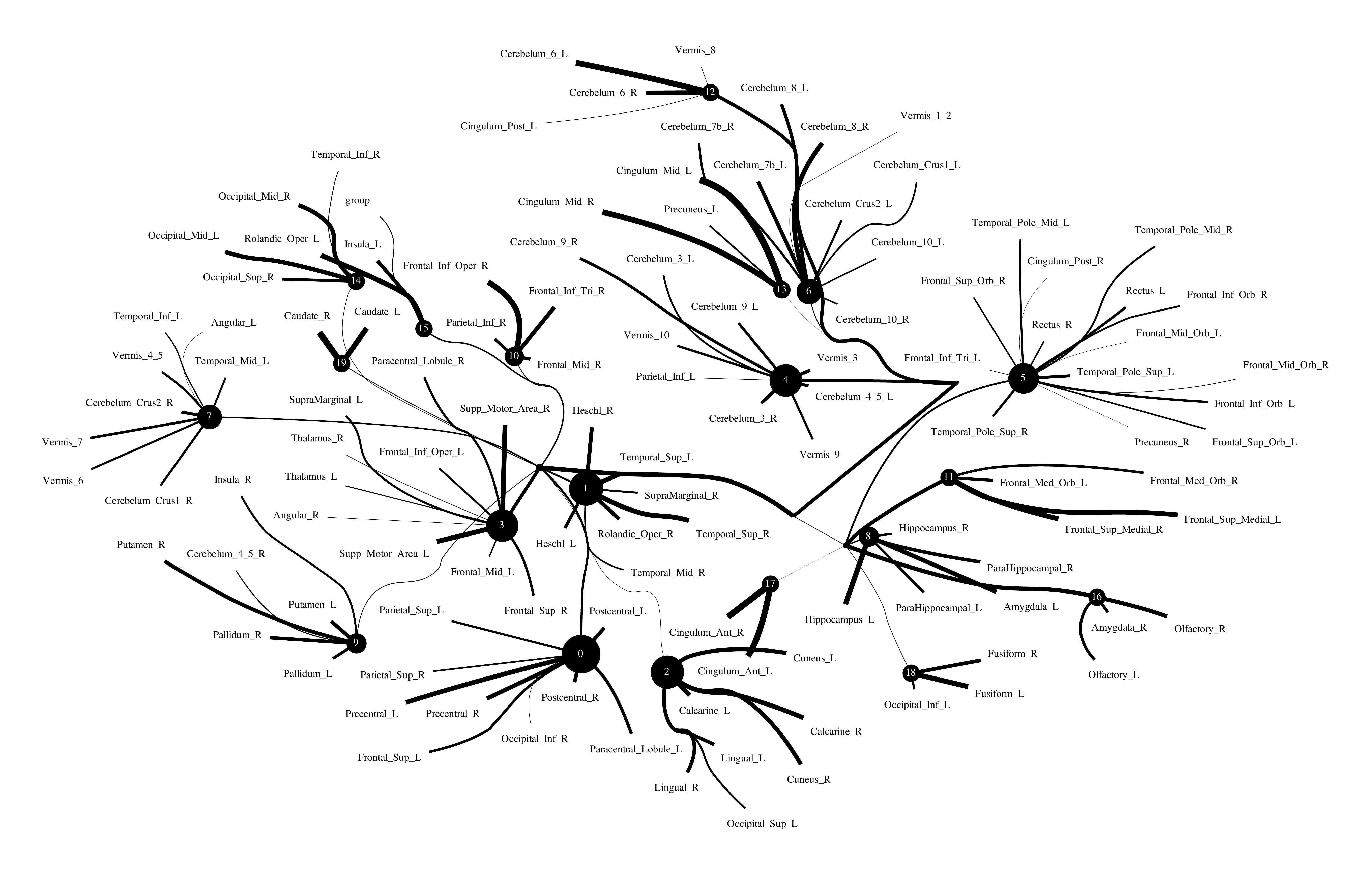}}}
    \caption{\textbf{Functional connectivity of health groups.}} 
\end{sidewaysfigure}

\newpage
\begin{sidewaysfigure}
    \centering
    {\scalebox{1}{\includegraphics[width=\textwidth,height=50em]{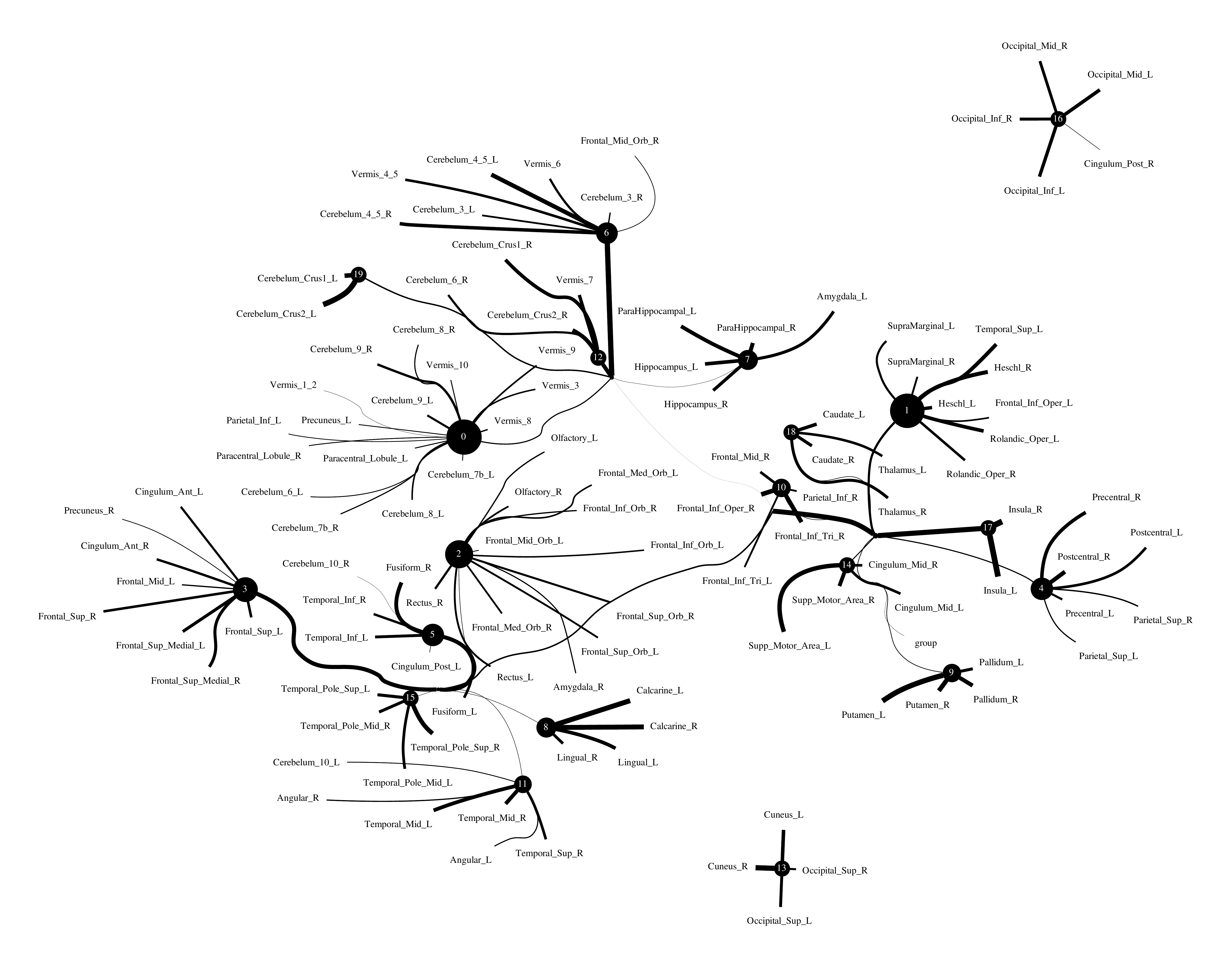}}}
    \caption{\textbf{Functional connectivity of patient groups.}} 
\end{sidewaysfigure}

\end{document}